\documentclass[aps,prl,showpacs,twocolumn,amsmath,amsfonts,amssymb,superscriptaddress,tightenlines]{revtex4-1}

\usepackage[colorlinks,citecolor=blue,linkcolor=cyan]{hyperref}
\usepackage{grffile}
\usepackage[caption=false]{subfig}
\usepackage{graphicx}
\usepackage{mathtools}

\renewcommand{\paragraph}[1]{{\par\it #1---}\ignorespaces}
\usepackage{xcolor}
\usepackage{paralist}
\newcommand{\dg}{^{\dagger}}
\newcommand{\ndg}{^{\phantom{}}}
\renewcommand{\vec}[1]{{\boldsymbol{#1}}}
\newcommand{\op}{\hat}
\newcommand{\id}{\mathbb{I}}
\newcommand{\wt}{\widetilde}
\newcommand{\Tr}{\textsf{Tr}}

\newcommand{\pr}{^{\prime}}

\newcommand{\erfc}{\text{erfc}}
\newcommand{\ceil}[1]{\lceil {#1} \rceil}
\newcommand{\dimp}{{D_{\mathcal{P}}}}
\newcommand{\dimc}{{D_{\tiny{\text{cutoff}}}}}
\newcommand{\dimf}{{D_{\mathcal{H}}}}
\newcommand{\dimm}{{d_{\mathcal{M}}}}
\newcommand{\hp}{\mathcal{H}_{\mathcal{P}}}
\newcommand{\stwo}{S_2}
\newcommand{\param}{\lambda}
\newcommand{\psim}{\Psi[\mathcal{M}]}

\begin{document}
\date{\today}

\title{Complexity and geometry of quantum state manifolds}

\author{Zhoushen Huang} \affiliation{Institute for Materials Science,
  Los Alamos National Laboratory, Los Alamos, NM 87545, USA}
\email{zsh@lanl.gov}
\author{Alexander V. Balatsky} \affiliation{Institute for Materials
  Science, Los Alamos National Laboratory, Los Alamos, NM 87545, USA}
\affiliation{NORDITA, Roslagstullsbacken 23, SE-106 91\ \ Stockholm,
  Sweden}
\affiliation{Department of Physics, University of Connecticut, Storrs, CT 06269, USA}
\email{avb@nordita.org}
\begin{abstract}
  We show that the Hilbert space spanned by a continuously
  parametrized wavefunction family---\emph{i.e.}, a quantum state
  manifold---is dominated by a subspace, onto which all member states
  have close to unity projection weight. Its characteristic
  dimensionality $\dimp$ is much smaller than the full Hilbert space
  dimension, and is equivalent to a statistical complexity measure
  $e^{S_2}$, where $S_2$ is the $2^{nd}$ Renyi entropy of the
  manifold. In the thermodynamic limit, $\dimp$ closely approximates
  the quantum geometric volume of the manifold under the Fubini-Study
  metric, revealing an intriguing connection between information and
  geometry. This connection persists in compact manifolds such as a
  twisted boundary phase, where the corresponding geometric
  circumference is lower bounded by a term proportional to its
  topological index, reminiscent of entanglement entropy.
\end{abstract}

\maketitle

\paragraph{Introduction}
A ubiquitous notion in quantum physics is a state manifold,
\emph{i.e.}, a continuously parametrized wavefunction family
$\psim \equiv \bigl\{ |\Psi(\vec\param)\rangle \in \mathcal{H} \bigl|
\vec\param \in \mathcal{M} \bigr\}$.  One example, typically
encountered in the study of quantum phase transitions
\cite{SachdevQPT, DuttaQPT}, is the ground state of a Hamiltonian
$H(\vec \param)$, where $\vec \param$ may represent external field,
interaction strength, adiabatic pumping, \emph{etc}.  Another one is
the Hilbert space trajectory generated by time evolution, \emph{e.g.},
after a quantum quench.  Conventionally, $\psim$ is analyzed through a
few carefully designed characteristic quantities such as order
parameters, topological indices, \emph{etc.}, which are highly system
dependent.  With the recent influx of machine learning techniques
\cite{Carrasquilla16, Wang16, Carleo16, Chng16, Liu16, Zhang16,
  Chen17, Deng17}, a new paradigm is emerging, where the task of
identifying such reductions is delegated to domain-agnostic
algorithms, and the role of human becomes instead to provide domain
specific data and interpretations.  In light of these developments, an
important question to understand is the fundamental information limit
of $\psim$, namely: How much data, in principle, is enough (to enable
a machine, say) to capture everything about $\psim$?  Standard quantum
mechanics posits that everything is encoded in wavefunctions, yet
their number in $\psim$ is nominally infinite. A more practical
formulation is thus: is there a ``compression'' of $\psim$, from which
all member wavefunctions can be efficiently reconstructed to a high
fidelity?

That such a scheme should indeed exist follows from an intuitive
observation: The fraction of the Hilbert space $\mathcal{H}$
``occupied'' by $\psim$ must be very small. This is because in a
manybody setting, the dimension of $\mathcal{H}$ is exponentially
large in system size, thus a random state in $\mathcal{H}$ has almost
zero overlap with any $|\Psi(\vec\param)\rangle$, and represents a
dimension that is irrelevant to $\psim$. Eliminating all such
dimensions brings out a much more compact subspace
$\hp \subset \mathcal{H}$ which remains representative of $\psim$,
thereby achieving a ``compression rate'' $\dimp / \dimf$, where
$\dimp$ is the characteristic dimensionality of $\hp$, and
$\dimf = \dim(\mathcal{H})$ is the full Hilbert space dimension.  We
will discuss how to identify $\hp$, and show that $\dimp$ is related
to the 2nd Renyi entropy $\stwo$ of $\psim$, as well as its quantum
geometric volume $\mathcal{V}$ under the Fubini-Study metric
\cite{Provost80},
\begin{gather}
  \label{dsv}
  \dimp = e^{\stwo} \simeq \mathcal{V} / \sqrt{\pi^{\dimm}}\ ,
\end{gather}
where $\dimm = \dim(\mathcal{M})$ is the parameter space dimension.
Toward the end we will also examplify a topological obstruction to the
trivialization of these quantities on a compact
manifold. Eq.~\ref{dsv} reveals an intriguing connection between
quantumness, information, and geometry, and is reminiscent of various
aspects of the ongoing discussion regarding the nature of their mutual
relations \cite{tHooft93, Hardy01, RyuTakayanagi06, Raamsdonk10,
  Swingle12, Qi13,*Lee15, Cao16}.

\paragraph{Principal vectors of a state manifold}
To characterize the Hilbert space structure of the state manifold
$\psim$, it is useful to think of it as a statistical ensemble.  The
eigenstates of its density operator $\op \rho$ then provide a basis
set of $\mathcal{H}$ which are naturally ranked by their relevance to
$\psim$, encoded as the eigenvalues.  We briefly discuss the eigen
decomposition of $\op \rho$ through the \emph{singular value
  decomposition} (SVD) of its amplitude matrix $A$.  Consider an
$M$-point uniform discretization of the parameter space
$\mathcal{M} \rightarrow \{\vec\param_1, \vec\param_2, \cdots,
\vec\param_M\}$.  Continuum limit can be restored at the end if
necessary.  Taking each
$|\Psi_m\rangle \equiv |\Psi(\vec\param_m)\rangle$ as a
$\dimf$-dimensional column vector, the amplitude matrix $A$ is
constructed as
\begin{gather}
  \label{adef}
  A = \frac{1}{\sqrt{M}}(|\Psi_1 \rangle\ |\Psi_2\rangle \ \cdots\ |\Psi_M\rangle) \ .
\end{gather}
SVD of $A$ then gives (subscripts denote matrix sizes)
\begin{gather}
  \label{svd}
  A_{\dimf \times M}\ndg = U_{\dimf \times K}\ndg\ \Lambda_{K \times K}\ndg\ V\dg_{\,K\times M}\ ,
\end{gather}
in which
$\Lambda_{\kappa\kappa\pr} =
\sqrt{w_{\kappa}}\delta_{\kappa,\kappa\pr}$ is a positive diagonal
matrix with $1 \ge w_1 \ge w_2 \ge \cdots w_K > 0$, $U$ and $V$ are
column unitary matrices, $U\dg U = V\dg V = \id_{K\times K}$, and
$K \le \min(\dimf, M)$ is the rank of $A$, which is also the
\emph{exact} dimensionality of the Hilbert space spanned by the $M$
sampled states.  In a manybody setting, $M \ll \dimf$, thus $K = M$ if
we assume all sampled states are linearly independent.  The eigen
decomposition of $\op \rho$ then reads
\begin{gather}
  \op \rho = AA\dg = \frac{1}{M} \sum_{m=1}^M |\Psi_m\rangle\langle
  \Psi_m| = \sum_{\kappa = 1}^K w_\kappa |U_\kappa\rangle\langle U_\kappa|\ ,
\end{gather}
where $|U_\kappa\rangle$ is the $\kappa^{th}$ column of $U$.  Note
that $\Tr (\op \rho) = \sum_{\kappa}w_{\kappa} = 1$.  The expansion
coefficient
$\langle U_\kappa | \Psi_m\rangle = (U\dg A)_{\kappa m} = \sqrt{M
  w_{\kappa}} V\dg_{\kappa m}$ only involves $V$ and $\{w_{\kappa}\}$,
and can be obtained alternatively via the eigen decomposition of the
$M \times M$ overlap matrix $G$,
\begin{gather}
  \label{g-def}
  G_{mn} \equiv \frac{1}{M}\langle \Psi_m | \Psi_n\rangle = (A\dg A)_{mn}\quad , \quad G = V\dg
  \Lambda V\ ,
\end{gather}
which will be useful in numerical schemes where overlaps can be
efficiently computed.  In statistical sciences, $\{|U_\kappa\rangle\}$
are known as \emph{principal vectors}, and the above procedure amounts
to the \emph{principal component analysis} (PCA) of
$\{|\Psi_m\rangle\}$. PCA of spin configuration data has recently been
used to identify phase transitions \cite{Wang16, Wetzel17, Hu17}.

\begin{figure}
  \centering
  \subfloat[$180$ evenly spaced samples]{
    \includegraphics[width=.23\textwidth, trim={20 0 0
      0},clip]
    {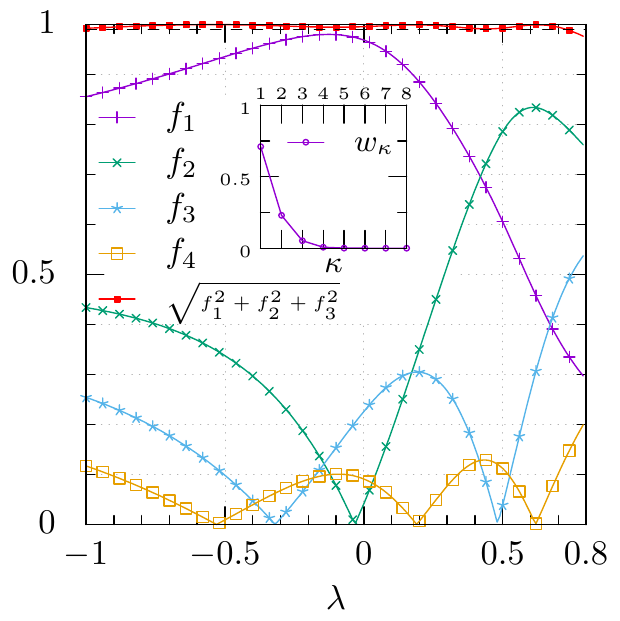}
  }
  \subfloat[$4$ samples at $\lambda = \pm 0.25, \pm 0.75$]{
    \hspace{-1em}\includegraphics[width=.26\textwidth]
    {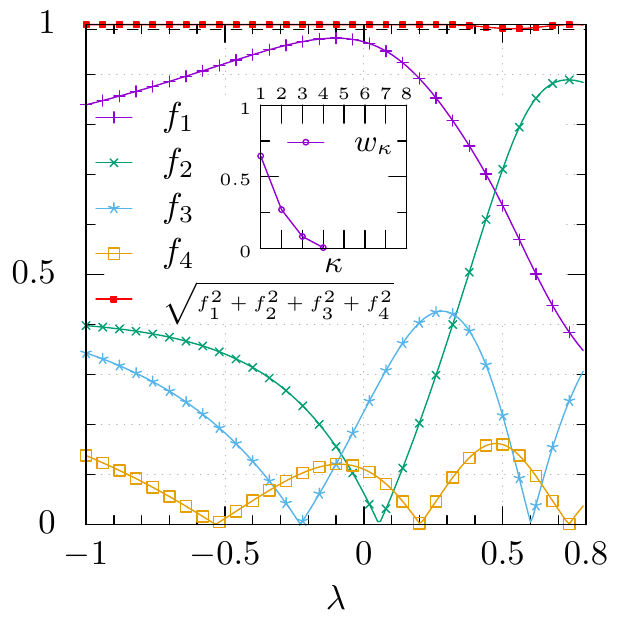}
  }
  \caption{PCA of the bilinear-biquadratic model with $14$ spins,
    using (a) $M=180$ ground states evenly sampled between
    $-1 < \lambda < 0.8$, and (b) $M=4$ samples at
    $\lambda = \pm 0.25, \pm 0.75$.
    $f_{\kappa} = | \langle U_k | \Psi(\lambda) \rangle |$ are the
    expansion coefficients of the exact states in the principal vector
    basis ($1 \le \kappa \le M$). Insets show the first few principal
    weights. Note that in both cases, (i) the first three weights
    nearly exhaust the full weight, $w_1 + w_2 + w_3 \simeq 0.99$, and
    (ii) The $\kappa^{th}$ expansion coefficient $f_{\kappa}$ has
    $\kappa - 1$ nodes. Red squared lines show projection
    \emph{amplitude} of the exact states onto a $3$ or $4$ dimensional
    principal Hilbert space, which almost always exceeds $0.99$ (black
    dashed line).}
  \label{fig:bil-biq}
\end{figure}

We illustrate the representational power of the principal vectors
using the bilinear-biquadratic $S=1$ spin chain \cite{Lauchli06},
$H(\param) = \sum_{x=1}^N \vec S_x\cdot \vec S_{x+1} + \param (\vec
S_x \cdot \vec S_{x+1})^2$, where $N$ is chain length, and
$\vec S_{N+1} \equiv \vec S_{1}$ (periodic boundary condition).  For
$N=14$, we compute its ground state (exact diagonalization) at
$M = 180$ points uniformly sampled between $-1 < \lambda < 0.8$, which
covers most of the gapped Haldane phase \cite{Haldane83, AKLT87}
\footnote{We set the upper limit to $0.8$ because for the finite chain
  length $N=14$ used here, a premature transition to the gapless phase
  occurs at $\lambda$ slightly above $0.8$, due to finite size
  effect. At the thermodynamic limit $N \rightarrow \infty$, this
  transition should happen at $\lambda = 1$. }. Their PCA, shown in
Fig.~\ref{fig:bil-biq}(a), has several interesting features:

(i) The $\kappa^{th}$ expansion coefficient
$f_\kappa(\param) \equiv \langle U_\kappa |\Psi(\param)\rangle$ has
$\kappa-1$ nodes in the $\lambda$ space. This is reminiscent of the
radial nodal structure in the hydrogen atom, which is attributed to
the orthogonality between states of different principal quanutm
numbers. A similar orthogonality exists between $f_\kappa$ and
$f_{\kappa\pr}$:
$\int d\lambda\, f_\kappa(\lambda)f_{\kappa\pr}^{*}(\lambda) / \int
d\lambda \rightarrow \frac{1}{M}\sum_{m = 1}^M f_\kappa(\lambda_m)
f^{*}_{k\pr}(\lambda_m) = (\Lambda V\dg V \Lambda)_{\kappa\kappa\pr} =
w_\kappa \delta_{\kappa,\kappa\pr}$, which follows from Eq.~\ref{svd}
\footnote{ In fact, all $f_k(\lambda)$ are real for the
  bilinear-biquadratic model, which follows the real-valuedness of the
  Hamiltonian (and hence its ground state wavefunction). Orthogonality
  does not guarantee complex wavefunctions to have node(s) along the
  $\lambda$ axis. }, and is suggestive of an ``emergent quantum
mechanics'' in the $\param$ space.

(ii) The first few principal vectors already reproduce $\sim 1$ total
weight. For the $N=14$ chain, the leading weights ($w_{\kappa}$) are
$\{0.71, 0.23, 0.052, 0.0067, \cdots\}$, thus a cutoff dimension at
$\dimc = 3$, for example, preserves a total weight of $\sim
0.99$. Note that $\dimc \ll \dimf = 3^{14}$. Indeed, one can define a
\emph{principal Hilbert space} $\hp$ as the span of the first $\dimc$
principal vectors; then as shown in Fig.~\ref{fig:bil-biq}(a), most
ground states within the sampled region can be approximated to a
fidelity of $\gtrsim 0.99$ by its projection onto $\hp$, \emph{i.e.},
a $3$-term truncation
$|\Psi(\lambda)\rangle \simeq \sum_{\kappa=1}^3 f_{\kappa}(\lambda)
|U_{\kappa}\rangle$. 

The compactness of $\hp$ suggests that a sparse sampling of the
$\param$ space at $M \sim \dimc$ points could gather enough
information to generate as good a set of principal vectors as a dense
sampling.  In Fig.~\ref{fig:bil-biq}(b), we show that taking $M=4$
samples is sufficient to approximate any \emph{unsampled} state in the
manifold. Thus the entire ground state manifold can be approximately
reconstructed by diagonalizing a $4\times 4$ Hamiltonian
$\wt H_{\kappa\kappa\pr}(\param) = \langle U_\kappa | H(\param) |
U_{\kappa\pr}\rangle$ for \emph{continuous} $\param$, using
$\{|U_\kappa\rangle\}$ generated from a $4$-point sampling. Similarly,
all physical operators are approximately $4\times 4$ matrices.

\paragraph{Principal dimensionality} The impressive compactness of
$\hp$ prompts the question: what determines an appropriate cutoff
dimension? Below we provide an estimation.  Consider a state
$|\Psi_m\rangle \in \psim$. Its average projection weight onto a
random state in $\psim$ is
\begin{gather}
  \label{wm-def}
  \wt w_m \equiv \frac{1}{M} \sum_{n=1}^M \Bigl| \langle \Psi_m | \Psi_n\rangle \Bigr|^2\ ,
\end{gather}
thus it takes approximately $D_m = \lceil \frac{1}{\wt w_m} \rceil$
random states in $\psim$ to supply a total weight of $\sim 1$. In the
thermodynamic limit, these $D_m$ random states are themselves mutually
orthogonal, and can serve as a minimal basis set to expand
$|\Psi_m\rangle$. A further average over $m$ removes the
$m$-dependence, and there are two natural choices:
$1/\langle \wt w_m\rangle$ and $\langle 1/\wt w_m\rangle$. We adopt
the first one as it puts the $m$ and $n$ indices on an equal footing,
but remark that the second option would eliminate a dominant
correction ($\langle \sigma^2\rangle$) from Eq.~\ref{dp-l} and
therefore has its own merit \footnote{See Supplemental Materials for
  additional details.}. This leads to our definition of
\emph{principal dimensionality},
\begin{gather}
  \label{dp-def}
  \dimp \equiv \frac{1}{\langle \wt w_m\rangle} = \frac{M^2}{\sum_{m,n=1}^{M} \Bigl|\langle \Psi_m |
    \Psi_n \rangle \Bigr|^2} = \frac{1}{\Tr(G^2)}\ ,
\end{gather}
where
$\langle \cdots \rangle = \sum_m(\cdots) /M \stackrel{M \rightarrow \infty}{\longrightarrow} \int d\vec
\param\, (\cdots) / \int d\vec \param$.

The last equality in Eq.~\ref{dp-def} suggests an interesting
connection with Renyi entropy. Recall that the $\alpha^{th}$ Renyi
entropy of a trace-normalized hermitian $\op{\mathcal{O}}$ is
$S_{\alpha}[\op{\mathcal{O}}] = \frac{1}{1-\alpha}\log
\Tr(\op{\mathcal{O}}^{\alpha})$, thus $\dimp = e^{S_2[G]}$. Since
$S_{\alpha}[\op{\mathcal{O}}]$ can be evaluated using the nonzero
eigenvalues of $\op{\mathcal{O}}$, two operators with identical
nonzero spectra---in this case $\op \rho = AA\dg$ and
$G = A\dg A$---must have the same Renyi entropy, hence
\begin{gather}
  \label{dp-es2}
  \dimp = e^{S_2[G]} = e^{S_2[\op \rho]} = \frac{1}{\sum_{\kappa=1}^K w_{\kappa}^2}\ .
\end{gather}
Our definition of $\dimp$, based on considerations of wavefunction
overlaps, is thus consistent with the intuitive correspondence between
number of effective degrees of freedom and exponential of entropies
\cite{LandauLifshitzV}.  The physical meaning of $\dimp$ is now more
transparent: the last expression is an \emph{inverse participation
  ratio}, thus $\dimp$ represents the number of principal vectors
participating in the manifold $\psim$, and is therefore a natural
measure of its effective Hilbert space dimension, and equivalently,
its degree of \emph{Hilbert space localization} \cite{Cohen16}.

Note that $\dimp$ thus defined is not an integer.  For numerical
comparison, we introduce an integer truncation dimension
$D(W)\equiv \min_\kappa \left[\sum_{\kappa\pr = 1}^\kappa
  w_{\kappa\pr} \ge W\right]$, \emph{i.e.}, the smallest number of
principal vectors needed to reach a total weight of $W$ .  We expect
$\dimp$ to be comparable to a $D(W)$ with $W \sim 1$.

Since exact diagonalization of interacting models, such as the
bilinear-biquadratic spin chain used earlier, can only be implemented
for small system size, we now switch to a free fermion model to
illustrate $\dimp$ at large $N$. The Su-Schreiffer-Heeger (SSH) model
\cite{SSH79} describes fermions hopping on a one-dimensional lattice
with two alternating hopping amplitudes ($1$ and $\lambda$) on
neighboring bonds. With $2N$ lattice sites, its Hamiltonian is
$H(\lambda) = \sum_{x=1}^N c_{2x-1}\dg c_{2x} + \param c_{2x}\dg
c_{2x+1} + h.c.$, and we use periodic boundary condition
$(c_{2N+1}, c_{2N+2}) = (c_1, c_2)$. Its ground state at half filling
is
$|\Psi(\param)\rangle = \prod_{a} \psi\dg_{k_a}(\param) |
\emptyset\rangle$, where $\psi_{k_a}\dg(\param)$ creates a lower band
eigenmode of momentum $k_a = 2\pi a/N$ for $a = 1, 2, \cdots, N$.  The
weights $\{w_{\kappa}\}$ can be obtained by diagonalizing the overlap
matrix
$G_{\lambda \lambda\pr} = \langle \Psi(\lambda) |
\Psi(\lambda\pr)\rangle = \prod_a\langle \emptyset |
\psi_{k_a}(\lambda) \psi_{k_a}\dg(\lambda\pr) | \emptyset\rangle$.

\begin{figure}
  \centering
  \includegraphics[width=.45\textwidth]
  {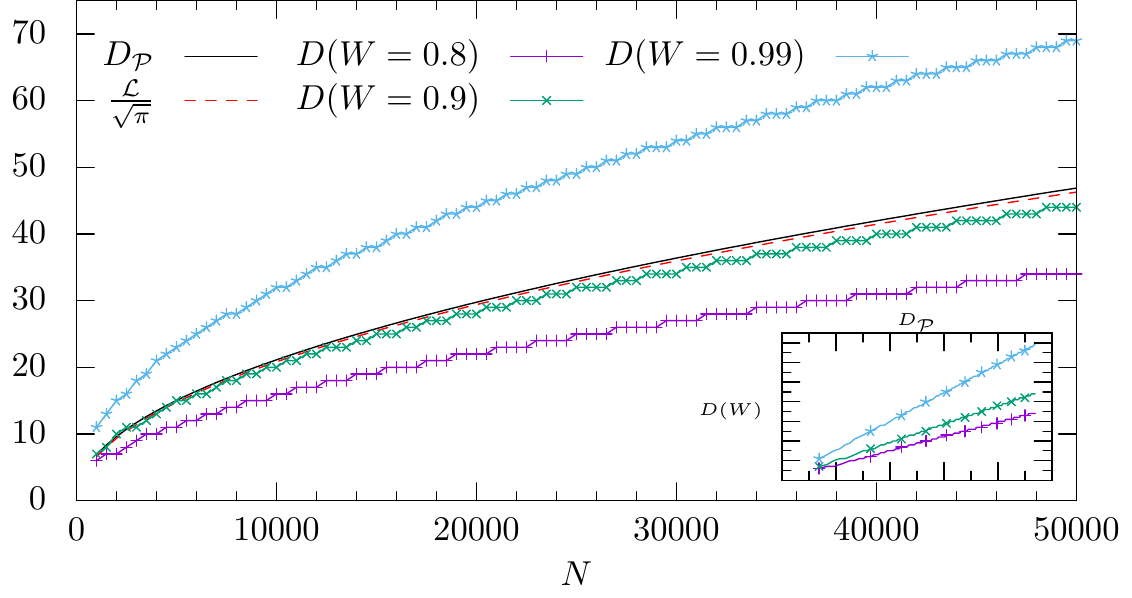}
  \caption{Comparison of principal dimensionality $\dimp$, geometric
    length $\mathcal{L}$, and truncation dimensions $D(W)$ at
    different weights $W$, using the SSH model with $2N$ lattice sites
    and $1000$ ground states evenly sampled between
    $-0.5 < \lambda < 0.5$. Note that for all $N$,
    $\dimp\simeq \frac{\mathcal{L}}{\sqrt{\pi}}$, and exceeds
    $D(0.9)$, \emph{i.e.}, the first $\ceil{\dimp}$ principal vectors
    represent $> 90\%$ total weight. Inset shows that all $D(W)$ are
    linearly related to $\dimp$.}
  \label{fig:ssh-pbc}
\end{figure}

In Fig.~\ref{fig:ssh-pbc}, we uniformly sample $M = 1000$ ground
states within $-0.5 < \param < 0.5$, and plot $\dimp$ and $D(W)$ for
several $W$'s, as functions of system size $N$. As shown,
$\dimp > D(0.9)$, \emph{i.e.}, the first $\lceil \dimp \rceil$
principal vectors represent over $90\%$ total weight of the state
manifold. In addition, $D(W)\forall W$ are linearly related to $\dimp$
(see inset).  Thus $\dimp$ accurately captures the Hilbert space size
of the SSH ground state manifold.

\paragraph{Relation with quantum geometry}
The geometric content of quantum state manifolds has been intensively
studied in the past decade, particularly in relation with quantum
phases and phase transitions \cite{Carollo05, Zhu06, Zanardi06,
  Zanardi07, You07, Venuti07, Chen08, Matsuura10, Ma13, Neupert13,
  Kolodrubetz13, Kolodrubetz16}.  Since two wavefunctions that differ
by an infinitesimal $\delta \vec \param$ are in general not
orthogonal, their overlap amplitude can be interpreted as a Hilbert
space distance,
$d(\Psi, \Phi) = \cos^{-1} \bigl|\langle \Psi | \Phi\rangle \bigr|$, a
measure of their maximal experimental distinguishability
\cite{Wootters81}. This endows $\psim$ with a metric \cite{Zanardi07},
$F_{\mu\nu}(\vec \param) \equiv \langle
\partial_{\param_{\mu}}\Psi(\vec \param) | \partial_{\param_{\nu}}
\Psi(\vec \param)\rangle - \langle \partial_{\param_{\mu}}\Psi(\vec
\param) | \Psi(\vec \param) \rangle\langle \Psi(\vec \param) |
\partial_{\param_{\nu}} \Psi(\vec \param)\rangle$, which turns out to
encode surprisingly rich physics.  Its anti-symmetric and symmetric
parts are, respectively, the Berry curvature
$\Omega_{\mu\nu} = -i(F_{\mu\nu} - F_{\nu\mu})$ \cite{Berry84}, and
the Fubini-Study metric
$g_{\mu\nu}= \frac{1}{2}(F_{\mu\nu}+F_{\nu\mu})$ \cite{Provost80}.
The former has long been known to play a fundamental role in systems
with nontrivial topology \cite{TKNN82, Berry84}. Study of the latter
in the context of quantum manybody physics initiated only more
recently with Refs.~\cite{Zanardi07, Venuti07}, which showed that
$F_{\mu\nu}$ exhibits universal scaling near phase transitions, and
becomes singular at quantum critical points.  The more global aspects
of $F_{\mu\nu}$ have since been systematically investigated
\cite{Matsuura10, Ma13, Kolodrubetz13, Kolodrubetz16}.  Note that the
Fubini-Study metric can be viewed as a quantum generalization of the
Fisher-Rao information metric \cite{Braunstein94}, and the divergence
of the latter has been shown to signify \emph{thermal} phase
transitions \cite{Prokopenko11}, \emph{vis-a-vis} its quantum
counterpart.

The appearance of wavefunction overlaps in Eq.~\ref{dp-def} suggests a
potential geometric interpretation of $\dimp$ and $S_2$.
Interestingly, Hall \cite{Hall99} has argued from general grounds that
$e^{S_1}$---where $S_1$ is the von Neumann entropy---can be viewed as
a geometric volume measure for any statistical ensemble, although
Renyi entropies $S_{\alpha}$ with $\alpha \neq 0, 1$ were explicitly
excluded. We now establish a connection between $\dimp$ and the
Fubini-Study volume.

For clarity, we will consider a scalar parameter
$\param$. Straightforward Taylor expansion gives
$\bigl| \langle \Psi_{\lambda} | \Psi_{\lambda+\Delta \lambda}\rangle
\bigr|^2 = 1 - g_{\lambda} \Delta\lambda^2 +
\mathcal{O}(\Delta\lambda^3)$, where $g_{\lambda}$ is the Fubini-Study
metric,
\begin{gather}
  \label{fs-metric-scalar}
  g(\lambda) = \langle \partial_{\lambda} \Psi |
  \partial_{\lambda}\Psi \rangle - \langle \partial_{\lambda} \Psi |
  \Psi\rangle\langle \Psi | \partial_{\lambda}\Psi\rangle\ ,
\end{gather}
note that its length element reproduces the Hilbert space distance,
$\delta \ell = \sqrt{g(\lambda)}\,\delta\lambda = \cos^{-1}\bigl|
\langle \Psi_{\lambda} | \Psi_{\lambda + \delta\lambda}\rangle
\bigr|$, and is invariant under reparametrization. In the
thermodynamic limit, the overlap quickly drops to zero as
$\Delta \lambda$ increases, thus one can write
\begin{gather}
  \label{overlap-gaussian}
  \bigl| \langle \Psi(\param)| \Psi(\param + \Delta \param)\rangle \bigr|^2 \simeq e^{-g(\param) \Delta\param^2}\ .
\end{gather}
For a parameter space $\lambda \in (\lambda_{a}, \lambda_{b})$, this
allows us to rewrite Eq.~\ref{wm-def}, in the continuum limit, as
\begin{align}
  \label{w-cont}
  \wt w(\lambda) = \int\limits_{\lambda_a}^{\lambda_b} d\lambda\pr\, \frac{e^{- g(\lambda) (\lambda\pr - \lambda)^2}}{\lambda_b-\lambda_a}
  = 
  \sqrt{\frac{\pi}{g(\lambda)}}\,\frac{1-\xi(\lambda)}{\lambda_b-\lambda_a}\ ,
\end{align}
where $\xi(\lambda)$ accounts for the effect of finite integration
limits,
$\xi(\lambda) = \frac{1}{2}\erfc\bigl[\sqrt{g(\lambda)}(\lambda -
\lambda_a) \bigr] + \frac{1}{2}
\erfc\bigl[\sqrt{g(\lambda)}(\lambda_b-\lambda)\bigr]$, and
$\erfc(x) = \frac{2}{\sqrt{\pi}}\int_x^{\infty}dt\, e^{-t^2}$.
$\xi(\lambda) \simeq \frac{1}{2}$ at the boundaries $\lambda_{a,b}$,
but rapidly approaches zero away from them \cite{Note3}. As discussed
before, $1\over \wt w(\lambda)$ counts the number of random states in
$\psim$ needed to span $|\Psi(\lambda)\rangle$, thus away from the
boundaries $\lambda_{a,b}$, $\sqrt{g}$ is a ``local density of
dimensions'',
\begin{gather}
  \label{ldos}
  \sqrt{\frac{g(\lambda)}{\pi}}\simeq
  \frac{\wt w(\lambda)^{-1}}{\lambda_b - \lambda_a} = \parbox{9em}{\centering Density of \\principal dimension}\ .
\end{gather}
One can now anticipate that the quantum geometric length
$\mathcal{L}=\int d\lambda\, \sqrt{g(\lambda)}$, an integrated
``dimension density'', should recover the principal
dimensionality. Indeed, we have \cite{Note3}
\begin{gather}
  \label{dp-l}
  \dimp = \frac{\mathcal{L}}{\sqrt{\pi}} (1 + c)\quad , \quad \mathcal{L} \equiv \int\limits_{\lambda_a}^{\lambda_b} d\lambda\, \sqrt{g(\lambda)} \ ,
\end{gather}
where
$c = \langle \frac{1 - \xi(\lambda)}{1 + \sigma(\lambda)}\rangle^{-1}
- 1 = \langle \xi(\lambda)\rangle - \langle \sigma(\lambda)^2\rangle +
\cdots$,
$\sigma(\lambda) = \frac{\sqrt{g(\lambda)}}{\langle
  \sqrt{g(\lambda)}\rangle} - 1$, and
$\langle \cdots \rangle = \frac{\int \cdots \, d\lambda}{\int
  d\lambda}$.  $\langle\sigma^2\rangle$ is the relative standard
deviation of $\sqrt{g}$ independent of system size $N$.
$\langle \xi \rangle$ is an ``edge'' effect around $\lambda_{a,b}$,
and decreases with increasing $N$. Thus for a smooth enough $\psim$,
$c$ should be negligible.  In Fig.~\ref{fig:ssh-pbc}, we show that for
the SSH ground state manifold, $\mathcal{L}/\sqrt{\pi}$ almost
perfectly tracks $\dimp$.  The Gaussian integral Eq.~\ref{w-cont} can
easily accommodate $\dimm$-dimensional parameters, with which the
dominant contribution in Eq.~\ref{dp-l} becomes
$\dimp = \mathcal{V}/\sqrt{\pi^{\dimm}}$, where
$\mathcal{V} = \int d^{\dimm}\vec \lambda\, \sqrt{\det(g_{\mu\nu}(\vec
  \lambda))}$ is the Fubini-Study volume. Note that for gapped ground
state manifolds, $g_{\mu\nu}\propto N$ where $N$ is system size
\cite{Venuti07}, thus $\dimp \propto N^{\dimm/2}$.

\paragraph{Compact manifold and topology}
Compact manifolds may host topologically protected degeneracies, which
can be interpreted as a lower bound to Hilbert space dimension. This
in turn implies topological lower bounds to $S_2$ and
$\mathcal{L}$. We briefly discuss this point using again the SSH
model, but on the compact space of a twisted boundary phase
$\theta \in [0, 2\pi]$ \cite{Niu85,Hirano08},
$c_{2N+i} = e^{i\theta} c_i$, $i=1,2$. The ground state at half
filling becomes
$|\Psi(\lambda, \theta)\rangle = \prod_a\psi_{k_a(\theta)}\dg(\lambda)
| \emptyset\rangle$, where $k_a(\theta) = \frac{2\pi a + \theta}{N}$,
$a = 1, 2, \cdots, N$. Since Bloch states with different $\theta$ are
no longer orthogonal, the overlap matrix elements become Slater
determinants,
$G_{\theta \theta\pr}(\lambda) = \langle \Psi(\lambda,\theta) |
\Psi(\lambda, \theta\pr)\rangle = \det \mathcal{G}(\lambda)$, where
$\mathcal{G}_{ab}(\lambda) = \langle \emptyset |
\psi_{k_a(\theta)}(\lambda)
\psi_{k_b(\theta\pr)}\dg(\lambda)|\emptyset\rangle$. The system
undergoes a topological phase transition at $|\lambda_c| = 1$, where
its Berry phase
$\gamma = -\int_0^{2\pi}d\theta \langle \Psi |
i\partial_{\theta}\Psi\rangle$ changes discretely from $0$
($|\lambda| < |\lambda_c|$) to $\pi$ ($|\lambda| > |\lambda_c|$).

\begin{figure}
  \centering
  \subfloat[$4$ leading principal weights]{
    \includegraphics[width=.23\textwidth]
    {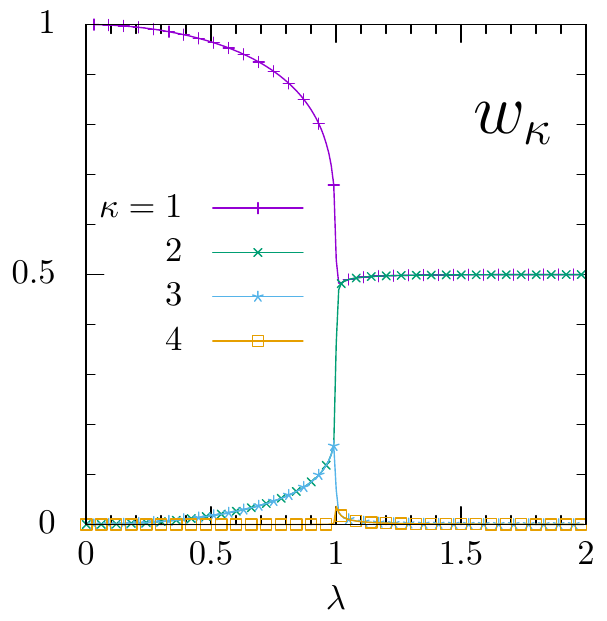}
  }
  \subfloat[$\dimp$ vs $\mathcal{L}$]{
    \includegraphics[width=.22\textwidth]
    {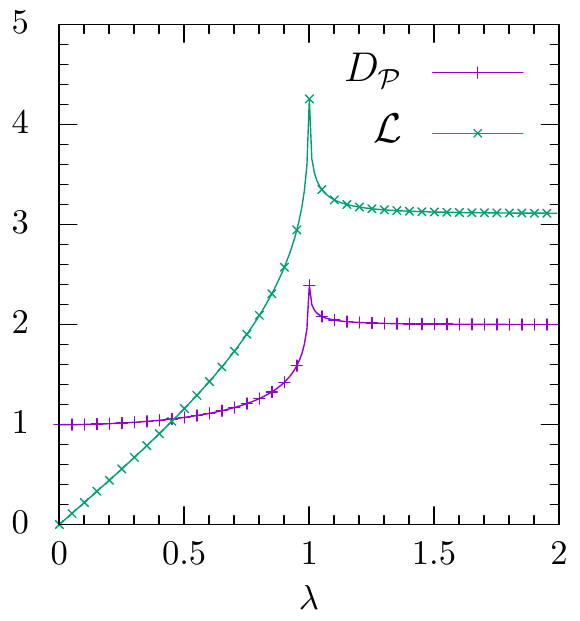}
  }
  \caption{SSH model with twisted boundary phase
    $\theta \in [0,2\pi]$, on $2N=400$ lattice sites. Its topological
    index is $|\nu| = \Theta(|\lambda| - 1)$. (a) shows the first four
    weights. There is one dominant weight in the non-topological phase
    ($\lambda < 1$), and two in the topological phase ($\lambda >
    1$). (b) $\dimp$ is consistent with level counting in (a), and
    $\mathcal{L}$ is lower bounded by $|\nu|\pi$, see text for
    discussion.  }
  \label{fig:ssh-tbc}
\end{figure}

In Fig.~\ref{fig:ssh-tbc}, we take a $2N=400$-site chain and compute
$\{w_{\kappa}\}$, $\dimp$, and $\mathcal{L}$ for its ground state
ensemble $\{|\Psi(\theta | \lambda)\rangle\}$ at 100 $\theta$s evenly
sampled between $0$ and $2\pi$, and plot them as functions of
$\lambda \in [0,2]$. In the non-topological phase, a single weight
dominates, with two degenerate subleading ones appreciating near the
critical $\lambda_c = 1$; $\dimp$ stays close to $1$ and only exhibits
a sharp upturn near $\lambda_c$. The topological phase is dominated by
two degenerate weights, thus $\dimp$ stabilizes toward $2$. The cusp
at $\lambda_c$ reflects the topological phase transition.

The behavior of $\mathcal{L}$ mirrors that of $\dimp$, but does not
\emph{quantitatively} satisfy $\dimp \simeq \mathcal{L}/\sqrt{\pi}$.
This is expected, because the deviation of
$\langle \Psi(\theta) | \Psi(\theta + \Delta \theta)\rangle$ from $1$
is driven solely by the \emph{single} bond on which the twisted
boundary phase is applied, which thus spoils the manybody character of
Eq.~\ref{overlap-gaussian} and hence its quantitative accuracy.  To
acquire some analytical understanding, we truncate to a
two-dimensional subspace, justified because $\dimp \le 2$ in both
phases except near the critical point. One can then write (dropping
$\lambda$ dependence)
$|\Psi(\theta)\rangle = \frac{1}{\sqrt{2}} \left[|\wt U_1\rangle +
  e^{i\phi(\theta)} |\wt U_2\rangle \right]$, which is the most
general form of a two-component wavefunction that can host a robust
winding number,
$\nu = \int_0^{2\pi} \frac{d\theta}{2\pi} \partial_{\theta}\phi$, and
$\gamma = \nu\pi$ is the Berry phase.  From
Eq.~\ref{fs-metric-scalar},
$g(\theta) = \frac{1}{4}(\partial_{\theta}\phi)^2$, thus
$\mathcal{L} = \frac{1}{2} \int_0^{2\pi} d\theta \left|
  \partial_{\theta}\phi \right| \ge \frac{1}{2} \left|\int_0^{2\pi}
  d\theta \, \partial_{\theta}\phi \right| = |\nu| \pi$. Indeed, in
Fig.~\ref{fig:ssh-tbc}(b), $\mathcal{L}$ stabilizes to $\pi$ for
$\lambda > \lambda_c$ and touches $0$ at $\lambda = 0$.  The geometric
length of a compact manifold is thus lower bounded by its topological
index, reminiscent of the behavior of entanglement entropy
\cite{RyuHatsugai06, CalabreseCardy04}.

\paragraph{Conclusion and discussions}
We have shown that the effective Hilbert space dimension $\dimp$ of a
quantum state manifold, its complexity $e^{S_2}$, and its Fubini-Study
geometric volume $\mathcal{V}$ are quantitatively related,
$\dimp = e^{S_2}\simeq \mathcal{V} / \sqrt{\pi^{\dimm}}$. On a compact
manifold, there is a topological obstruction to their trivialization,
with a lower bound determined by the topological index.  The
interpretation of geometric volume as a Hilbert space dimension is
suggestive of an approximate ``geometric quantization'', in unit of
$\sqrt{\pi^{\dimm}}$, of parameter spaces (which may coincide with
real space and time). This is analogous to the quantization of
classical phase space in unit of $(2\pi \hbar)^d$. In this sense,
$\dimp$ is a natural quantum generalization of the notion of phase
space volume. For a $\vec\lambda$ that parametrizes a ground state,
$\dimp$ represents the amount of quantum fluctuation driven by
$\vec \lambda$. Divergence of $g_{\mu\nu}$ at critical point
\cite{Venuti07} then implies that as one approaches quantum
criticality, increasingly more states are ``pulled down'' from higher
energies to span the effective Hilbert space.  For a manifold
generated from unitary time evolution (i.e., $\lambda = $ time),
$\ceil{\dimp}$ is roughly the number of Hilbert space dimensions
``activated'' over the course of time, and can be measured as the
integrated energy fluctuation \cite{Anandan90}. In the context of
adiabatic quantum computation, this becomes the number of orthogonal
``machine states'' needed to carry out an algorithm, and is therefore
a kind of computational complexity.

\begin{acknowledgments}
  \paragraph{Acknowledgments}
  We are grateful to J.-X.~Zhu and W.~Zhu for critical reading and
  comments on an early draft, and to D.~P.~Arovas, G.~Aeppli, W.~Zhu,
  J.-X.~Zhu, A.~Saxena, H.~Choi, T.~Ahmed, and F.~Ronning for various
  discussions. Work at LANL was supported by US DOE NNSA through LANL
  LDRD (XWNK). Work at NORDITA was supported by ERC DM 321031.
\end{acknowledgments}
\bibliography{phs}

\begin{thebibliography}{53}%
\makeatletter
\providecommand \@ifxundefined [1]{%
 \@ifx{#1\undefined}
}%
\providecommand \@ifnum [1]{%
 \ifnum #1\expandafter \@firstoftwo
 \else \expandafter \@secondoftwo
 \fi
}%
\providecommand \@ifx [1]{%
 \ifx #1\expandafter \@firstoftwo
 \else \expandafter \@secondoftwo
 \fi
}%
\providecommand \natexlab [1]{#1}%
\providecommand \enquote  [1]{``#1''}%
\providecommand \bibnamefont  [1]{#1}%
\providecommand \bibfnamefont [1]{#1}%
\providecommand \citenamefont [1]{#1}%
\providecommand \href@noop [0]{\@secondoftwo}%
\providecommand \href [0]{\begingroup \@sanitize@url \@href}%
\providecommand \@href[1]{\@@startlink{#1}\@@href}%
\providecommand \@@href[1]{\endgroup#1\@@endlink}%
\providecommand \@sanitize@url [0]{\catcode `\\12\catcode `\$12\catcode
  `\&12\catcode `\#12\catcode `\^12\catcode `\_12\catcode `\%12\relax}%
\providecommand \@@startlink[1]{}%
\providecommand \@@endlink[0]{}%
\providecommand \url  [0]{\begingroup\@sanitize@url \@url }%
\providecommand \@url [1]{\endgroup\@href {#1}{\urlprefix }}%
\providecommand \urlprefix  [0]{URL }%
\providecommand \Eprint [0]{\href }%
\providecommand \doibase [0]{http://dx.doi.org/}%
\providecommand \selectlanguage [0]{\@gobble}%
\providecommand \bibinfo  [0]{\@secondoftwo}%
\providecommand \bibfield  [0]{\@secondoftwo}%
\providecommand \translation [1]{[#1]}%
\providecommand \BibitemOpen [0]{}%
\providecommand \bibitemStop [0]{}%
\providecommand \bibitemNoStop [0]{.\EOS\space}%
\providecommand \EOS [0]{\spacefactor3000\relax}%
\providecommand \BibitemShut  [1]{\csname bibitem#1\endcsname}%
\let\auto@bib@innerbib\@empty
\bibitem [{\citenamefont {Sachdev}(2001)}]{SachdevQPT}%
  \BibitemOpen
  \bibfield  {author} {\bibinfo {author} {\bibfnamefont {S.}~\bibnamefont
  {Sachdev}},\ }\href@noop {} {\emph {\bibinfo {title} {Quantum Phase
  Transitions}}}\ (\bibinfo  {publisher} {Cambridge University Press},\
  \bibinfo {year} {2001})\BibitemShut {NoStop}%
\bibitem [{\citenamefont {{Dutta}}\ \emph {et~al.}(2010)\citenamefont
  {{Dutta}}, \citenamefont {{Aeppli}}, \citenamefont {{Chakrabarti}},
  \citenamefont {{Divakaran}}, \citenamefont {{Rosenbaum}},\ and\ \citenamefont
  {{Sen}}}]{DuttaQPT}%
  \BibitemOpen
  \bibfield  {author} {\bibinfo {author} {\bibfnamefont {A.}~\bibnamefont
  {{Dutta}}}, \bibinfo {author} {\bibfnamefont {G.}~\bibnamefont {{Aeppli}}},
  \bibinfo {author} {\bibfnamefont {B.~K.}\ \bibnamefont {{Chakrabarti}}},
  \bibinfo {author} {\bibfnamefont {U.}~\bibnamefont {{Divakaran}}}, \bibinfo
  {author} {\bibfnamefont {T.~F.}\ \bibnamefont {{Rosenbaum}}}, \ and\ \bibinfo
  {author} {\bibfnamefont {D.}~\bibnamefont {{Sen}}},\ }\href@noop {} {\emph
  {\bibinfo {title} {{Quantum phase transitions in transverse field spin
  models: from statistical physics to quantum information}}}}\ (\bibinfo {year}
  {2010})\ \Eprint {http://arxiv.org/abs/1012.0653} {arXiv:1012.0653
  [cond-mat.stat-mech]} \BibitemShut {NoStop}%
\bibitem [{\citenamefont {{Carrasquilla}}\ and\ \citenamefont
  {{Melko}}(2017)}]{Carrasquilla16}%
  \BibitemOpen
  \bibfield  {author} {\bibinfo {author} {\bibfnamefont {J.}~\bibnamefont
  {{Carrasquilla}}}\ and\ \bibinfo {author} {\bibfnamefont {R.~G.}\
  \bibnamefont {{Melko}}},\ }\href {\doibase 10.1038/nphys4035} {\bibfield
  {journal} {\bibinfo  {journal} {Nature Physics}\ }\textbf {\bibinfo {volume}
  {13}},\ \bibinfo {pages} {431} (\bibinfo {year} {2017})},\ \Eprint
  {http://arxiv.org/abs/1605.01735} {arXiv:1605.01735 [cond-mat.str-el]}
  \BibitemShut {NoStop}%
\bibitem [{\citenamefont {{Wang}}(2016)}]{Wang16}%
  \BibitemOpen
  \bibfield  {author} {\bibinfo {author} {\bibfnamefont {L.}~\bibnamefont
  {{Wang}}},\ }\href {\doibase 10.1103/PhysRevB.94.195105} {\bibfield
  {journal} {\bibinfo  {journal} {\prb}\ }\textbf {\bibinfo {volume} {94}},\
  \bibinfo {eid} {195105} (\bibinfo {year} {2016})},\ \Eprint
  {http://arxiv.org/abs/1606.00318} {arXiv:1606.00318 [cond-mat.stat-mech]}
  \BibitemShut {NoStop}%
\bibitem [{\citenamefont {{Carleo}}\ and\ \citenamefont
  {{Troyer}}(2017)}]{Carleo16}%
  \BibitemOpen
  \bibfield  {author} {\bibinfo {author} {\bibfnamefont {G.}~\bibnamefont
  {{Carleo}}}\ and\ \bibinfo {author} {\bibfnamefont {M.}~\bibnamefont
  {{Troyer}}},\ }\href {\doibase 10.1126/science.aag2302} {\bibfield  {journal}
  {\bibinfo  {journal} {Science}\ }\textbf {\bibinfo {volume} {355}},\ \bibinfo
  {pages} {602} (\bibinfo {year} {2017})},\ \Eprint
  {http://arxiv.org/abs/1606.02318} {arXiv:1606.02318 [cond-mat.dis-nn]}
  \BibitemShut {NoStop}%
\bibitem [{\citenamefont {{Ch'ng}}\ \emph {et~al.}(2017)\citenamefont
  {{Ch'ng}}, \citenamefont {{Carrasquilla}}, \citenamefont {{Melko}},\ and\
  \citenamefont {{Khatami}}}]{Chng16}%
  \BibitemOpen
  \bibfield  {author} {\bibinfo {author} {\bibfnamefont {K.}~\bibnamefont
  {{Ch'ng}}}, \bibinfo {author} {\bibfnamefont {J.}~\bibnamefont
  {{Carrasquilla}}}, \bibinfo {author} {\bibfnamefont {R.~G.}\ \bibnamefont
  {{Melko}}}, \ and\ \bibinfo {author} {\bibfnamefont {E.}~\bibnamefont
  {{Khatami}}},\ }\href {\doibase 10.1103/PhysRevX.7.031038} {\bibfield
  {journal} {\bibinfo  {journal} {Physical Review X}\ }\textbf {\bibinfo
  {volume} {7}},\ \bibinfo {eid} {031038} (\bibinfo {year} {2017})},\ \Eprint
  {http://arxiv.org/abs/1609.02552} {arXiv:1609.02552 [cond-mat.str-el]}
  \BibitemShut {NoStop}%
\bibitem [{\citenamefont {{Liu}}\ \emph {et~al.}(2017)\citenamefont {{Liu}},
  \citenamefont {{Qi}}, \citenamefont {{Meng}},\ and\ \citenamefont
  {{Fu}}}]{Liu16}%
  \BibitemOpen
  \bibfield  {author} {\bibinfo {author} {\bibfnamefont {J.}~\bibnamefont
  {{Liu}}}, \bibinfo {author} {\bibfnamefont {Y.}~\bibnamefont {{Qi}}},
  \bibinfo {author} {\bibfnamefont {Z.~Y.}\ \bibnamefont {{Meng}}}, \ and\
  \bibinfo {author} {\bibfnamefont {L.}~\bibnamefont {{Fu}}},\ }\href {\doibase
  10.1103/PhysRevB.95.041101} {\bibfield  {journal} {\bibinfo  {journal}
  {\prb}\ }\textbf {\bibinfo {volume} {95}},\ \bibinfo {eid} {041101} (\bibinfo
  {year} {2017})},\ \Eprint {http://arxiv.org/abs/1610.03137} {arXiv:1610.03137
  [cond-mat.str-el]} \BibitemShut {NoStop}%
\bibitem [{\citenamefont {{Zhang}}\ and\ \citenamefont
  {{Kim}}(2017)}]{Zhang16}%
  \BibitemOpen
  \bibfield  {author} {\bibinfo {author} {\bibfnamefont {Y.}~\bibnamefont
  {{Zhang}}}\ and\ \bibinfo {author} {\bibfnamefont {E.-A.}\ \bibnamefont
  {{Kim}}},\ }\href {\doibase 10.1103/PhysRevLett.118.216401} {\bibfield
  {journal} {\bibinfo  {journal} {Physical Review Letters}\ }\textbf {\bibinfo
  {volume} {118}},\ \bibinfo {eid} {216401} (\bibinfo {year} {2017})},\ \Eprint
  {http://arxiv.org/abs/1611.01518} {arXiv:1611.01518 [cond-mat.str-el]}
  \BibitemShut {NoStop}%
\bibitem [{\citenamefont {{Chen}}\ \emph {et~al.}(2017)\citenamefont {{Chen}},
  \citenamefont {{Cheng}}, \citenamefont {{Xie}}, \citenamefont {{Wang}},\ and\
  \citenamefont {{Xiang}}}]{Chen17}%
  \BibitemOpen
  \bibfield  {author} {\bibinfo {author} {\bibfnamefont {J.}~\bibnamefont
  {{Chen}}}, \bibinfo {author} {\bibfnamefont {S.}~\bibnamefont {{Cheng}}},
  \bibinfo {author} {\bibfnamefont {H.}~\bibnamefont {{Xie}}}, \bibinfo
  {author} {\bibfnamefont {L.}~\bibnamefont {{Wang}}}, \ and\ \bibinfo {author}
  {\bibfnamefont {T.}~\bibnamefont {{Xiang}}},\ }\href@noop {} {\bibfield
  {journal} {\bibinfo  {journal} {ArXiv e-prints}\ } (\bibinfo {year}
  {2017})},\ \Eprint {http://arxiv.org/abs/1701.04831} {arXiv:1701.04831
  [cond-mat.str-el]} \BibitemShut {NoStop}%
\bibitem [{\citenamefont {{Deng}}\ \emph {et~al.}(2017)\citenamefont {{Deng}},
  \citenamefont {{Li}},\ and\ \citenamefont {{Das Sarma}}}]{Deng17}%
  \BibitemOpen
  \bibfield  {author} {\bibinfo {author} {\bibfnamefont {D.-L.}\ \bibnamefont
  {{Deng}}}, \bibinfo {author} {\bibfnamefont {X.}~\bibnamefont {{Li}}}, \ and\
  \bibinfo {author} {\bibfnamefont {S.}~\bibnamefont {{Das Sarma}}},\ }\href
  {\doibase 10.1103/PhysRevX.7.021021} {\bibfield  {journal} {\bibinfo
  {journal} {Physical Review X}\ }\textbf {\bibinfo {volume} {7}},\ \bibinfo
  {eid} {021021} (\bibinfo {year} {2017})},\ \Eprint
  {http://arxiv.org/abs/1701.04844} {arXiv:1701.04844 [cond-mat.dis-nn]}
  \BibitemShut {NoStop}%
\bibitem [{\citenamefont {Provost}\ and\ \citenamefont
  {Vallee}(1980)}]{Provost80}%
  \BibitemOpen
  \bibfield  {author} {\bibinfo {author} {\bibfnamefont {J.~P.}\ \bibnamefont
  {Provost}}\ and\ \bibinfo {author} {\bibfnamefont {G.}~\bibnamefont
  {Vallee}},\ }\href {https://projecteuclid.org:443/euclid.cmp/1103908308}
  {\bibfield  {journal} {\bibinfo  {journal} {Comm. Math. Phys.}\ }\textbf
  {\bibinfo {volume} {76}},\ \bibinfo {pages} {289} (\bibinfo {year}
  {1980})}\BibitemShut {NoStop}%
\bibitem [{\citenamefont {{'t Hooft}}(1993)}]{tHooft93}%
  \BibitemOpen
  \bibfield  {author} {\bibinfo {author} {\bibfnamefont {G.}~\bibnamefont {{'t
  Hooft}}},\ }\href@noop {} {\bibfield  {journal} {\bibinfo  {journal} {ArXiv
  General Relativity and Quantum Cosmology e-prints}\ } (\bibinfo {year}
  {1993})},\ \Eprint {http://arxiv.org/abs/gr-qc/9310026} {gr-qc/9310026}
  \BibitemShut {NoStop}%
\bibitem [{\citenamefont {{Hardy}}(2001)}]{Hardy01}%
  \BibitemOpen
  \bibfield  {author} {\bibinfo {author} {\bibfnamefont {L.}~\bibnamefont
  {{Hardy}}},\ }\href@noop {} {\bibfield  {journal} {\bibinfo  {journal}
  {eprint arXiv:quant-ph/0101012}\ } (\bibinfo {year} {2001})}\BibitemShut
  {NoStop}%
\bibitem [{\citenamefont {{Ryu}}\ and\ \citenamefont
  {{Takayanagi}}(2006)}]{RyuTakayanagi06}%
  \BibitemOpen
  \bibfield  {author} {\bibinfo {author} {\bibfnamefont {S.}~\bibnamefont
  {{Ryu}}}\ and\ \bibinfo {author} {\bibfnamefont {T.}~\bibnamefont
  {{Takayanagi}}},\ }\href {\doibase 10.1103/PhysRevLett.96.181602} {\bibfield
  {journal} {\bibinfo  {journal} {Physical Review Letters}\ }\textbf {\bibinfo
  {volume} {96}},\ \bibinfo {eid} {181602} (\bibinfo {year} {2006})},\ \Eprint
  {http://arxiv.org/abs/hep-th/0603001} {hep-th/0603001} \BibitemShut {NoStop}%
\bibitem [{\citenamefont {{van Raamsdonk}}(2010)}]{Raamsdonk10}%
  \BibitemOpen
  \bibfield  {author} {\bibinfo {author} {\bibfnamefont {M.}~\bibnamefont {{van
  Raamsdonk}}},\ }\href {\doibase 10.1007/s10714-010-1034-0} {\bibfield
  {journal} {\bibinfo  {journal} {General Relativity and Gravitation}\ }\textbf
  {\bibinfo {volume} {42}},\ \bibinfo {pages} {2323} (\bibinfo {year}
  {2010})},\ \Eprint {http://arxiv.org/abs/1005.3035} {arXiv:1005.3035
  [hep-th]} \BibitemShut {NoStop}%
\bibitem [{\citenamefont {{Swingle}}(2012)}]{Swingle12}%
  \BibitemOpen
  \bibfield  {author} {\bibinfo {author} {\bibfnamefont {B.}~\bibnamefont
  {{Swingle}}},\ }\href {\doibase 10.1103/PhysRevD.86.065007} {\bibfield
  {journal} {\bibinfo  {journal} {\prd}\ }\textbf {\bibinfo {volume} {86}},\
  \bibinfo {eid} {065007} (\bibinfo {year} {2012})},\ \Eprint
  {http://arxiv.org/abs/0905.1317} {arXiv:0905.1317 [cond-mat.str-el]}
  \BibitemShut {NoStop}%
\bibitem [{\citenamefont {{Qi}}(2013)}]{Qi13}%
  \BibitemOpen
  \bibfield  {author} {\bibinfo {author} {\bibfnamefont {X.-L.}\ \bibnamefont
  {{Qi}}},\ }\href@noop {} {\bibfield  {journal} {\bibinfo  {journal} {ArXiv
  e-prints}\ } (\bibinfo {year} {2013})},\ \Eprint
  {http://arxiv.org/abs/1309.6282} {arXiv:1309.6282 [hep-th]} \BibitemShut
  {NoStop}%
\bibitem [{\citenamefont {{Lee}}\ and\ \citenamefont {{Qi}}(2016)}]{Lee15}%
  \BibitemOpen
  \bibfield  {author} {\bibinfo {author} {\bibfnamefont {C.~H.}\ \bibnamefont
  {{Lee}}}\ and\ \bibinfo {author} {\bibfnamefont {X.-L.}\ \bibnamefont
  {{Qi}}},\ }\href {\doibase 10.1103/PhysRevB.93.035112} {\bibfield  {journal}
  {\bibinfo  {journal} {\prb}\ }\textbf {\bibinfo {volume} {93}},\ \bibinfo
  {eid} {035112} (\bibinfo {year} {2016})},\ \Eprint
  {http://arxiv.org/abs/1503.08592} {arXiv:1503.08592 [hep-th]} \BibitemShut
  {NoStop}%
\bibitem [{\citenamefont {{Cao}}\ \emph {et~al.}(2017)\citenamefont {{Cao}},
  \citenamefont {{Carroll}},\ and\ \citenamefont {{Michalakis}}}]{Cao16}%
  \BibitemOpen
  \bibfield  {author} {\bibinfo {author} {\bibfnamefont {C.}~\bibnamefont
  {{Cao}}}, \bibinfo {author} {\bibfnamefont {S.~M.}\ \bibnamefont
  {{Carroll}}}, \ and\ \bibinfo {author} {\bibfnamefont {S.}~\bibnamefont
  {{Michalakis}}},\ }\href {\doibase 10.1103/PhysRevD.95.024031} {\bibfield
  {journal} {\bibinfo  {journal} {\prd}\ }\textbf {\bibinfo {volume} {95}},\
  \bibinfo {eid} {024031} (\bibinfo {year} {2017})},\ \Eprint
  {http://arxiv.org/abs/1606.08444} {arXiv:1606.08444 [hep-th]} \BibitemShut
  {NoStop}%
\bibitem [{\citenamefont {{Wetzel}}(2017)}]{Wetzel17}%
  \BibitemOpen
  \bibfield  {author} {\bibinfo {author} {\bibfnamefont {S.~J.}\ \bibnamefont
  {{Wetzel}}},\ }\href {\doibase 10.1103/PhysRevE.96.022140} {\bibfield
  {journal} {\bibinfo  {journal} {\pre}\ }\textbf {\bibinfo {volume} {96}},\
  \bibinfo {eid} {022140} (\bibinfo {year} {2017})},\ \Eprint
  {http://arxiv.org/abs/1703.02435} {arXiv:1703.02435 [cond-mat.stat-mech]}
  \BibitemShut {NoStop}%
\bibitem [{\citenamefont {{Hu}}\ \emph {et~al.}(2017)\citenamefont {{Hu}},
  \citenamefont {{Singh}},\ and\ \citenamefont {{Scalettar}}}]{Hu17}%
  \BibitemOpen
  \bibfield  {author} {\bibinfo {author} {\bibfnamefont {W.}~\bibnamefont
  {{Hu}}}, \bibinfo {author} {\bibfnamefont {R.~R.~P.}\ \bibnamefont
  {{Singh}}}, \ and\ \bibinfo {author} {\bibfnamefont {R.~T.}\ \bibnamefont
  {{Scalettar}}},\ }\href {\doibase 10.1103/PhysRevE.95.062122} {\bibfield
  {journal} {\bibinfo  {journal} {\pre}\ }\textbf {\bibinfo {volume} {95}},\
  \bibinfo {eid} {062122} (\bibinfo {year} {2017})},\ \Eprint
  {http://arxiv.org/abs/1704.00080} {arXiv:1704.00080 [cond-mat.stat-mech]}
  \BibitemShut {NoStop}%
\bibitem [{\citenamefont {L\"auchli}\ \emph {et~al.}(2006)\citenamefont
  {L\"auchli}, \citenamefont {Schmid},\ and\ \citenamefont
  {Trebst}}]{Lauchli06}%
  \BibitemOpen
  \bibfield  {author} {\bibinfo {author} {\bibfnamefont {A.}~\bibnamefont
  {L\"auchli}}, \bibinfo {author} {\bibfnamefont {G.}~\bibnamefont {Schmid}}, \
  and\ \bibinfo {author} {\bibfnamefont {S.}~\bibnamefont {Trebst}},\ }\href
  {\doibase 10.1103/PhysRevB.74.144426} {\bibfield  {journal} {\bibinfo
  {journal} {Phys. Rev. B}\ }\textbf {\bibinfo {volume} {74}},\ \bibinfo
  {pages} {144426} (\bibinfo {year} {2006})},\ \Eprint
  {http://arxiv.org/abs/cond-mat/0607173} {cond-mat/0607173} \BibitemShut
  {NoStop}%
\bibitem [{\citenamefont {Haldane}(1983)}]{Haldane83}%
  \BibitemOpen
  \bibfield  {author} {\bibinfo {author} {\bibfnamefont {F.~D.~M.}\
  \bibnamefont {Haldane}},\ }\href {\doibase 10.1103/PhysRevLett.50.1153}
  {\bibfield  {journal} {\bibinfo  {journal} {Phys. Rev. Lett.}\ }\textbf
  {\bibinfo {volume} {50}},\ \bibinfo {pages} {1153} (\bibinfo {year}
  {1983})}\BibitemShut {NoStop}%
\bibitem [{\citenamefont {Affleck}\ \emph {et~al.}(1987)\citenamefont
  {Affleck}, \citenamefont {Kennedy}, \citenamefont {Lieb},\ and\ \citenamefont
  {Tasaki}}]{AKLT87}%
  \BibitemOpen
  \bibfield  {author} {\bibinfo {author} {\bibfnamefont {I.}~\bibnamefont
  {Affleck}}, \bibinfo {author} {\bibfnamefont {T.}~\bibnamefont {Kennedy}},
  \bibinfo {author} {\bibfnamefont {E.~H.}\ \bibnamefont {Lieb}}, \ and\
  \bibinfo {author} {\bibfnamefont {H.}~\bibnamefont {Tasaki}},\ }\href
  {\doibase 10.1103/PhysRevLett.59.799} {\bibfield  {journal} {\bibinfo
  {journal} {Phys. Rev. Lett.}\ }\textbf {\bibinfo {volume} {59}},\ \bibinfo
  {pages} {799} (\bibinfo {year} {1987})}\BibitemShut {NoStop}%
\bibitem [{Note1()}]{Note1}%
  \BibitemOpen
  \bibinfo {note} {We set the upper limit to $0.8$ because for the finite chain
  length $N=14$ used here, a premature transition to the gapless phase occurs
  at $\lambda $ slightly above $0.8$, due to finite size effect. At the
  thermodynamic limit $N \rightarrow \infty $, this transition should happen at
  $\lambda = 1$.}\BibitemShut {Stop}%
\bibitem [{Note2()}]{Note2}%
  \BibitemOpen
  \bibinfo {note} {In fact, all $f_k(\lambda )$ are real for the
  bilinear-biquadratic model, which follows the real-valuedness of the
  Hamiltonian (and hence its ground state wavefunction). Orthogonality does not
  guarantee complex wavefunctions to have node(s) along the $\lambda $
  axis.}\BibitemShut {Stop}%
\bibitem [{Note3()}]{Note3}%
  \BibitemOpen
  \bibinfo {note} {See Supplemental Materials for additional
  details.}\BibitemShut {Stop}%
\bibitem [{\citenamefont {Landau}\ and\ \citenamefont
  {Lifshitz}(1969)}]{LandauLifshitzV}%
  \BibitemOpen
  \bibfield  {author} {\bibinfo {author} {\bibfnamefont {L.~D.}\ \bibnamefont
  {Landau}}\ and\ \bibinfo {author} {\bibfnamefont {E.~M.}\ \bibnamefont
  {Lifshitz}},\ }\href@noop {} {\emph {\bibinfo {title} {Statistical Physics:
  V. 5: Course of Theoretical Physics}}}\ (\bibinfo  {publisher} {Pergamon
  press},\ \bibinfo {year} {1969})\BibitemShut {NoStop}%
\bibitem [{\citenamefont {{Cohen}}\ \emph {et~al.}(2016)\citenamefont
  {{Cohen}}, \citenamefont {{Yukalov}},\ and\ \citenamefont
  {{Ziegler}}}]{Cohen16}%
  \BibitemOpen
  \bibfield  {author} {\bibinfo {author} {\bibfnamefont {D.}~\bibnamefont
  {{Cohen}}}, \bibinfo {author} {\bibfnamefont {V.~I.}\ \bibnamefont
  {{Yukalov}}}, \ and\ \bibinfo {author} {\bibfnamefont {K.}~\bibnamefont
  {{Ziegler}}},\ }\href {\doibase 10.1103/PhysRevA.93.042101} {\bibfield
  {journal} {\bibinfo  {journal} {\pra}\ }\textbf {\bibinfo {volume} {93}},\
  \bibinfo {eid} {042101} (\bibinfo {year} {2016})},\ \Eprint
  {http://arxiv.org/abs/1511.04667} {arXiv:1511.04667 [cond-mat.quant-gas]}
  \BibitemShut {NoStop}%
\bibitem [{\citenamefont {Su}\ \emph {et~al.}(1979)\citenamefont {Su},
  \citenamefont {Schrieffer},\ and\ \citenamefont {Heeger}}]{SSH79}%
  \BibitemOpen
  \bibfield  {author} {\bibinfo {author} {\bibfnamefont {W.~P.}\ \bibnamefont
  {Su}}, \bibinfo {author} {\bibfnamefont {J.~R.}\ \bibnamefont {Schrieffer}},
  \ and\ \bibinfo {author} {\bibfnamefont {A.~J.}\ \bibnamefont {Heeger}},\
  }\href {\doibase 10.1103/PhysRevLett.42.1698} {\bibfield  {journal} {\bibinfo
   {journal} {Phys. Rev. Lett.}\ }\textbf {\bibinfo {volume} {42}},\ \bibinfo
  {pages} {1698} (\bibinfo {year} {1979})}\BibitemShut {NoStop}%
\bibitem [{\citenamefont {{Carollo}}\ and\ \citenamefont
  {{Pachos}}(2005)}]{Carollo05}%
  \BibitemOpen
  \bibfield  {author} {\bibinfo {author} {\bibfnamefont {A.~C.~M.}\
  \bibnamefont {{Carollo}}}\ and\ \bibinfo {author} {\bibfnamefont {J.~K.}\
  \bibnamefont {{Pachos}}},\ }\href {\doibase 10.1103/PhysRevLett.95.157203}
  {\bibfield  {journal} {\bibinfo  {journal} {Physical Review Letters}\
  }\textbf {\bibinfo {volume} {95}},\ \bibinfo {eid} {157203} (\bibinfo {year}
  {2005})},\ \Eprint {http://arxiv.org/abs/cond-mat/0502272} {cond-mat/0502272}
  \BibitemShut {NoStop}%
\bibitem [{\citenamefont {{Zhu}}(2006)}]{Zhu06}%
  \BibitemOpen
  \bibfield  {author} {\bibinfo {author} {\bibfnamefont {S.-L.}\ \bibnamefont
  {{Zhu}}},\ }\href {\doibase 10.1103/PhysRevLett.96.077206} {\bibfield
  {journal} {\bibinfo  {journal} {Physical Review Letters}\ }\textbf {\bibinfo
  {volume} {96}},\ \bibinfo {eid} {077206} (\bibinfo {year} {2006})},\ \Eprint
  {http://arxiv.org/abs/cond-mat/0511565} {cond-mat/0511565} \BibitemShut
  {NoStop}%
\bibitem [{\citenamefont {{Zanardi}}\ and\ \citenamefont
  {{Paunkovi{\'c}}}(2006)}]{Zanardi06}%
  \BibitemOpen
  \bibfield  {author} {\bibinfo {author} {\bibfnamefont {P.}~\bibnamefont
  {{Zanardi}}}\ and\ \bibinfo {author} {\bibfnamefont {N.}~\bibnamefont
  {{Paunkovi{\'c}}}},\ }\href {\doibase 10.1103/PhysRevE.74.031123} {\bibfield
  {journal} {\bibinfo  {journal} {\pre}\ }\textbf {\bibinfo {volume} {74}},\
  \bibinfo {eid} {031123} (\bibinfo {year} {2006})},\ \Eprint
  {http://arxiv.org/abs/quant-ph/0512249} {quant-ph/0512249} \BibitemShut
  {NoStop}%
\bibitem [{\citenamefont {Zanardi}\ \emph {et~al.}(2007)\citenamefont
  {Zanardi}, \citenamefont {Giorda},\ and\ \citenamefont
  {Cozzini}}]{Zanardi07}%
  \BibitemOpen
  \bibfield  {author} {\bibinfo {author} {\bibfnamefont {P.}~\bibnamefont
  {Zanardi}}, \bibinfo {author} {\bibfnamefont {P.}~\bibnamefont {Giorda}}, \
  and\ \bibinfo {author} {\bibfnamefont {M.}~\bibnamefont {Cozzini}},\ }\href
  {\doibase 10.1103/PhysRevLett.99.100603} {\bibfield  {journal} {\bibinfo
  {journal} {Phys. Rev. Lett.}\ }\textbf {\bibinfo {volume} {99}},\ \bibinfo
  {pages} {100603} (\bibinfo {year} {2007})}\BibitemShut {NoStop}%
\bibitem [{\citenamefont {{You}}\ \emph {et~al.}(2007)\citenamefont {{You}},
  \citenamefont {{Li}},\ and\ \citenamefont {{Gu}}}]{You07}%
  \BibitemOpen
  \bibfield  {author} {\bibinfo {author} {\bibfnamefont {W.-L.}\ \bibnamefont
  {{You}}}, \bibinfo {author} {\bibfnamefont {Y.-W.}\ \bibnamefont {{Li}}}, \
  and\ \bibinfo {author} {\bibfnamefont {S.-J.}\ \bibnamefont {{Gu}}},\ }\href
  {\doibase 10.1103/PhysRevE.76.022101} {\bibfield  {journal} {\bibinfo
  {journal} {\pre}\ }\textbf {\bibinfo {volume} {76}},\ \bibinfo {eid} {022101}
  (\bibinfo {year} {2007})},\ \Eprint {http://arxiv.org/abs/quant-ph/0701077}
  {quant-ph/0701077} \BibitemShut {NoStop}%
\bibitem [{\citenamefont {{Campos Venuti}}\ and\ \citenamefont
  {{Zanardi}}(2007)}]{Venuti07}%
  \BibitemOpen
  \bibfield  {author} {\bibinfo {author} {\bibfnamefont {L.}~\bibnamefont
  {{Campos Venuti}}}\ and\ \bibinfo {author} {\bibfnamefont {P.}~\bibnamefont
  {{Zanardi}}},\ }\href {\doibase 10.1103/PhysRevLett.99.095701} {\bibfield
  {journal} {\bibinfo  {journal} {Physical Review Letters}\ }\textbf {\bibinfo
  {volume} {99}},\ \bibinfo {eid} {095701} (\bibinfo {year} {2007})},\ \Eprint
  {http://arxiv.org/abs/0705.2211} {arXiv:0705.2211 [quant-ph]} \BibitemShut
  {NoStop}%
\bibitem [{\citenamefont {{Chen}}\ \emph {et~al.}(2008)\citenamefont {{Chen}},
  \citenamefont {{Wang}}, \citenamefont {{Hao}},\ and\ \citenamefont
  {{Wang}}}]{Chen08}%
  \BibitemOpen
  \bibfield  {author} {\bibinfo {author} {\bibfnamefont {S.}~\bibnamefont
  {{Chen}}}, \bibinfo {author} {\bibfnamefont {L.}~\bibnamefont {{Wang}}},
  \bibinfo {author} {\bibfnamefont {Y.}~\bibnamefont {{Hao}}}, \ and\ \bibinfo
  {author} {\bibfnamefont {Y.}~\bibnamefont {{Wang}}},\ }\href {\doibase
  10.1103/PhysRevA.77.032111} {\bibfield  {journal} {\bibinfo  {journal}
  {\pra}\ }\textbf {\bibinfo {volume} {77}},\ \bibinfo {eid} {032111} (\bibinfo
  {year} {2008})},\ \Eprint {http://arxiv.org/abs/0801.0020} {arXiv:0801.0020}
  \BibitemShut {NoStop}%
\bibitem [{\citenamefont {{Matsuura}}\ and\ \citenamefont
  {{Ryu}}(2010)}]{Matsuura10}%
  \BibitemOpen
  \bibfield  {author} {\bibinfo {author} {\bibfnamefont {S.}~\bibnamefont
  {{Matsuura}}}\ and\ \bibinfo {author} {\bibfnamefont {S.}~\bibnamefont
  {{Ryu}}},\ }\href {\doibase 10.1103/PhysRevB.82.245113} {\bibfield  {journal}
  {\bibinfo  {journal} {\prb}\ }\textbf {\bibinfo {volume} {82}},\ \bibinfo
  {eid} {245113} (\bibinfo {year} {2010})},\ \Eprint
  {http://arxiv.org/abs/1007.2200} {arXiv:1007.2200 [cond-mat.mes-hall]}
  \BibitemShut {NoStop}%
\bibitem [{\citenamefont {{Ma}}\ \emph {et~al.}(2012)\citenamefont {{Ma}},
  \citenamefont {{Gu}}, \citenamefont {{Chen}}, \citenamefont {{Fan}},\ and\
  \citenamefont {{Liu}}}]{Ma13}%
  \BibitemOpen
  \bibfield  {author} {\bibinfo {author} {\bibfnamefont {Y.-Q.}\ \bibnamefont
  {{Ma}}}, \bibinfo {author} {\bibfnamefont {S.-J.}\ \bibnamefont {{Gu}}},
  \bibinfo {author} {\bibfnamefont {S.}~\bibnamefont {{Chen}}}, \bibinfo
  {author} {\bibfnamefont {H.}~\bibnamefont {{Fan}}}, \ and\ \bibinfo {author}
  {\bibfnamefont {W.-M.}\ \bibnamefont {{Liu}}},\ }\href@noop {} {\bibfield
  {journal} {\bibinfo  {journal} {ArXiv e-prints}\ } (\bibinfo {year}
  {2012})},\ \Eprint {http://arxiv.org/abs/1202.2397} {arXiv:1202.2397
  [cond-mat.str-el]} \BibitemShut {NoStop}%
\bibitem [{\citenamefont {{Neupert}}\ \emph {et~al.}(2013)\citenamefont
  {{Neupert}}, \citenamefont {{Chamon}},\ and\ \citenamefont
  {{Mudry}}}]{Neupert13}%
  \BibitemOpen
  \bibfield  {author} {\bibinfo {author} {\bibfnamefont {T.}~\bibnamefont
  {{Neupert}}}, \bibinfo {author} {\bibfnamefont {C.}~\bibnamefont {{Chamon}}},
  \ and\ \bibinfo {author} {\bibfnamefont {C.}~\bibnamefont {{Mudry}}},\ }\href
  {\doibase 10.1103/PhysRevB.87.245103} {\bibfield  {journal} {\bibinfo
  {journal} {\prb}\ }\textbf {\bibinfo {volume} {87}},\ \bibinfo {eid} {245103}
  (\bibinfo {year} {2013})},\ \Eprint {http://arxiv.org/abs/1303.4643}
  {arXiv:1303.4643 [cond-mat.str-el]} \BibitemShut {NoStop}%
\bibitem [{\citenamefont {{Kolodrubetz}}\ \emph {et~al.}(2013)\citenamefont
  {{Kolodrubetz}}, \citenamefont {{Gritsev}},\ and\ \citenamefont
  {{Polkovnikov}}}]{Kolodrubetz13}%
  \BibitemOpen
  \bibfield  {author} {\bibinfo {author} {\bibfnamefont {M.}~\bibnamefont
  {{Kolodrubetz}}}, \bibinfo {author} {\bibfnamefont {V.}~\bibnamefont
  {{Gritsev}}}, \ and\ \bibinfo {author} {\bibfnamefont {A.}~\bibnamefont
  {{Polkovnikov}}},\ }\href {\doibase 10.1103/PhysRevB.88.064304} {\bibfield
  {journal} {\bibinfo  {journal} {\prb}\ }\textbf {\bibinfo {volume} {88}},\
  \bibinfo {eid} {064304} (\bibinfo {year} {2013})},\ \Eprint
  {http://arxiv.org/abs/1305.0568} {arXiv:1305.0568 [cond-mat.stat-mech]}
  \BibitemShut {NoStop}%
\bibitem [{\citenamefont {{Kolodrubetz}}\ \emph {et~al.}(2017)\citenamefont
  {{Kolodrubetz}}, \citenamefont {{Sels}}, \citenamefont {{Mehta}},\ and\
  \citenamefont {{Polkovnikov}}}]{Kolodrubetz16}%
  \BibitemOpen
  \bibfield  {author} {\bibinfo {author} {\bibfnamefont {M.}~\bibnamefont
  {{Kolodrubetz}}}, \bibinfo {author} {\bibfnamefont {D.}~\bibnamefont
  {{Sels}}}, \bibinfo {author} {\bibfnamefont {P.}~\bibnamefont {{Mehta}}}, \
  and\ \bibinfo {author} {\bibfnamefont {A.}~\bibnamefont {{Polkovnikov}}},\
  }\href {\doibase 10.1016/j.physrep.2017.07.001} {\ \textbf {\bibinfo {volume}
  {697}},\ \bibinfo {pages} {1} (\bibinfo {year} {2017})},\ \Eprint
  {http://arxiv.org/abs/1602.01062} {arXiv:1602.01062 [cond-mat.quant-gas]}
  \BibitemShut {NoStop}%
\bibitem [{\citenamefont {Wootters}(1981)}]{Wootters81}%
  \BibitemOpen
  \bibfield  {author} {\bibinfo {author} {\bibfnamefont {W.~K.}\ \bibnamefont
  {Wootters}},\ }\href {\doibase 10.1103/PhysRevD.23.357} {\bibfield  {journal}
  {\bibinfo  {journal} {Phys. Rev. D}\ }\textbf {\bibinfo {volume} {23}},\
  \bibinfo {pages} {357} (\bibinfo {year} {1981})}\BibitemShut {NoStop}%
\bibitem [{\citenamefont {Berry}(1984)}]{Berry84}%
  \BibitemOpen
  \bibfield  {author} {\bibinfo {author} {\bibfnamefont {M.~V.}\ \bibnamefont
  {Berry}},\ }\href {\doibase 10.1098/rspa.1984.0023} {\bibfield  {journal}
  {\bibinfo  {journal} {Proceedings of the Royal Society of London A:
  Mathematical, Physical and Engineering Sciences}\ }\textbf {\bibinfo {volume}
  {392}},\ \bibinfo {pages} {45} (\bibinfo {year} {1984})}\BibitemShut
  {NoStop}%
\bibitem [{\citenamefont {Thouless}\ \emph {et~al.}(1982)\citenamefont
  {Thouless}, \citenamefont {Kohmoto}, \citenamefont {Nightingale},\ and\
  \citenamefont {den Nijs}}]{TKNN82}%
  \BibitemOpen
  \bibfield  {author} {\bibinfo {author} {\bibfnamefont {D.~J.}\ \bibnamefont
  {Thouless}}, \bibinfo {author} {\bibfnamefont {M.}~\bibnamefont {Kohmoto}},
  \bibinfo {author} {\bibfnamefont {M.~P.}\ \bibnamefont {Nightingale}}, \ and\
  \bibinfo {author} {\bibfnamefont {M.}~\bibnamefont {den Nijs}},\ }\href
  {\doibase 10.1103/PhysRevLett.49.405} {\bibfield  {journal} {\bibinfo
  {journal} {Phys. Rev. Lett.}\ }\textbf {\bibinfo {volume} {49}},\ \bibinfo
  {pages} {405} (\bibinfo {year} {1982})}\BibitemShut {NoStop}%
\bibitem [{\citenamefont {Braunstein}\ and\ \citenamefont
  {Caves}(1994)}]{Braunstein94}%
  \BibitemOpen
  \bibfield  {author} {\bibinfo {author} {\bibfnamefont {S.~L.}\ \bibnamefont
  {Braunstein}}\ and\ \bibinfo {author} {\bibfnamefont {C.~M.}\ \bibnamefont
  {Caves}},\ }\href {\doibase 10.1103/PhysRevLett.72.3439} {\bibfield
  {journal} {\bibinfo  {journal} {Phys. Rev. Lett.}\ }\textbf {\bibinfo
  {volume} {72}},\ \bibinfo {pages} {3439} (\bibinfo {year}
  {1994})}\BibitemShut {NoStop}%
\bibitem [{\citenamefont {Prokopenko}\ \emph {et~al.}(2011)\citenamefont
  {Prokopenko}, \citenamefont {Lizier}, \citenamefont {Obst},\ and\
  \citenamefont {Wang}}]{Prokopenko11}%
  \BibitemOpen
  \bibfield  {author} {\bibinfo {author} {\bibfnamefont {M.}~\bibnamefont
  {Prokopenko}}, \bibinfo {author} {\bibfnamefont {J.~T.}\ \bibnamefont
  {Lizier}}, \bibinfo {author} {\bibfnamefont {O.}~\bibnamefont {Obst}}, \ and\
  \bibinfo {author} {\bibfnamefont {X.~R.}\ \bibnamefont {Wang}},\ }\href
  {\doibase 10.1103/PhysRevE.84.041116} {\bibfield  {journal} {\bibinfo
  {journal} {Phys. Rev. E}\ }\textbf {\bibinfo {volume} {84}},\ \bibinfo
  {pages} {041116} (\bibinfo {year} {2011})}\BibitemShut {NoStop}%
\bibitem [{\citenamefont {{Hall}}(1999)}]{Hall99}%
  \BibitemOpen
  \bibfield  {author} {\bibinfo {author} {\bibfnamefont {M.~J.~W.}\
  \bibnamefont {{Hall}}},\ }\href {\doibase 10.1103/PhysRevA.59.2602}
  {\bibfield  {journal} {\bibinfo  {journal} {\pra}\ }\textbf {\bibinfo
  {volume} {59}},\ \bibinfo {pages} {2602} (\bibinfo {year} {1999})},\ \Eprint
  {http://arxiv.org/abs/physics/9903045} {physics/9903045} \BibitemShut
  {NoStop}%
\bibitem [{\citenamefont {Niu}\ \emph {et~al.}(1985)\citenamefont {Niu},
  \citenamefont {Thouless},\ and\ \citenamefont {Wu}}]{Niu85}%
  \BibitemOpen
  \bibfield  {author} {\bibinfo {author} {\bibfnamefont {Q.}~\bibnamefont
  {Niu}}, \bibinfo {author} {\bibfnamefont {D.~J.}\ \bibnamefont {Thouless}}, \
  and\ \bibinfo {author} {\bibfnamefont {Y.-S.}\ \bibnamefont {Wu}},\ }\href
  {\doibase 10.1103/PhysRevB.31.3372} {\bibfield  {journal} {\bibinfo
  {journal} {Phys. Rev. B}\ }\textbf {\bibinfo {volume} {31}},\ \bibinfo
  {pages} {3372} (\bibinfo {year} {1985})}\BibitemShut {NoStop}%
\bibitem [{\citenamefont {{Hirano}}\ \emph {et~al.}(2008)\citenamefont
  {{Hirano}}, \citenamefont {{Katsura}},\ and\ \citenamefont
  {{Hatsugai}}}]{Hirano08}%
  \BibitemOpen
  \bibfield  {author} {\bibinfo {author} {\bibfnamefont {T.}~\bibnamefont
  {{Hirano}}}, \bibinfo {author} {\bibfnamefont {H.}~\bibnamefont {{Katsura}}},
  \ and\ \bibinfo {author} {\bibfnamefont {Y.}~\bibnamefont {{Hatsugai}}},\
  }\href {\doibase 10.1103/PhysRevB.77.094431} {\bibfield  {journal} {\bibinfo
  {journal} {\prb}\ }\textbf {\bibinfo {volume} {77}},\ \bibinfo {eid} {094431}
  (\bibinfo {year} {2008})},\ \Eprint {http://arxiv.org/abs/0710.4198}
  {arXiv:0710.4198 [cond-mat.str-el]} \BibitemShut {NoStop}%
\bibitem [{\citenamefont {{Ryu}}\ and\ \citenamefont
  {{Hatsugai}}(2006)}]{RyuHatsugai06}%
  \BibitemOpen
  \bibfield  {author} {\bibinfo {author} {\bibfnamefont {S.}~\bibnamefont
  {{Ryu}}}\ and\ \bibinfo {author} {\bibfnamefont {Y.}~\bibnamefont
  {{Hatsugai}}},\ }\href {\doibase 10.1103/PhysRevB.73.245115} {\bibfield
  {journal} {\bibinfo  {journal} {\prb}\ }\textbf {\bibinfo {volume} {73}},\
  \bibinfo {eid} {245115} (\bibinfo {year} {2006})},\ \Eprint
  {http://arxiv.org/abs/cond-mat/0601237} {cond-mat/0601237} \BibitemShut
  {NoStop}%
\bibitem [{\citenamefont {{Calabrese}}\ and\ \citenamefont
  {{Cardy}}(2004)}]{CalabreseCardy04}%
  \BibitemOpen
  \bibfield  {author} {\bibinfo {author} {\bibfnamefont {P.}~\bibnamefont
  {{Calabrese}}}\ and\ \bibinfo {author} {\bibfnamefont {J.}~\bibnamefont
  {{Cardy}}},\ }\href {\doibase 10.1088/1742-5468/2004/06/P06002} {\bibfield
  {journal} {\bibinfo  {journal} {Journal of Statistical Mechanics: Theory and
  Experiment}\ }\textbf {\bibinfo {volume} {6}},\ \bibinfo {pages} {06002}
  (\bibinfo {year} {2004})},\ \Eprint {http://arxiv.org/abs/hep-th/0405152}
  {hep-th/0405152} \BibitemShut {NoStop}%
\bibitem [{\citenamefont {Anandan}\ and\ \citenamefont
  {Aharonov}(1990)}]{Anandan90}%
  \BibitemOpen
  \bibfield  {author} {\bibinfo {author} {\bibfnamefont {J.}~\bibnamefont
  {Anandan}}\ and\ \bibinfo {author} {\bibfnamefont {Y.}~\bibnamefont
  {Aharonov}},\ }\href {\doibase 10.1103/PhysRevLett.65.1697} {\bibfield
  {journal} {\bibinfo  {journal} {Phys. Rev. Lett.}\ }\textbf {\bibinfo
  {volume} {65}},\ \bibinfo {pages} {1697} (\bibinfo {year}
  {1990})}\BibitemShut {NoStop}%
\end{thebibliography}%

\onecolumngrid
\section{Supplemental Materials}
In this note, we provide derivation details and some numerical
justifications for Eq.(13) in the text, reproduced below,
\begin{gather}
  \dimp = \frac{\mathcal{L}}{\sqrt{\pi}}(1+c)\quad , \quad   \mathcal{L} = \int d\lambda \sqrt{g(\lambda)}
  \quad , \quad  c = \langle \frac{1-\xi(\lambda)}{1+\sigma(\lambda)}\rangle^{-1}\ .
\end{gather}
We also discuss an alternative definition of characteristic
dimensionality, $D_{alt} \equiv \langle 1/\wt w_m\rangle$ (whereas
$\dimp \equiv 1/\langle \wt w_m\rangle$).

Consider a one parameter wavefunction $|\Psi(\lambda)\rangle$, and
discretize the $\lambda$ space into a regular grid,
\begin{gather}
  \lambda_m = m \delta\lambda\quad , \quad m = 1, 2, \cdots, M\quad , \quad |\Psi_m\rangle \equiv |\Psi(\lambda_m)\rangle\ .
\end{gather}
Our task is to compute $\dimp$,
\begin{gather}
  \dimp = \sum_{m,n} \frac{M^2}{\left|\langle \Psi_m | \Psi_n\rangle\right|^2}\ .
\end{gather}

\begin{figure}
  \centering
  \includegraphics[width=.5\textwidth]
  {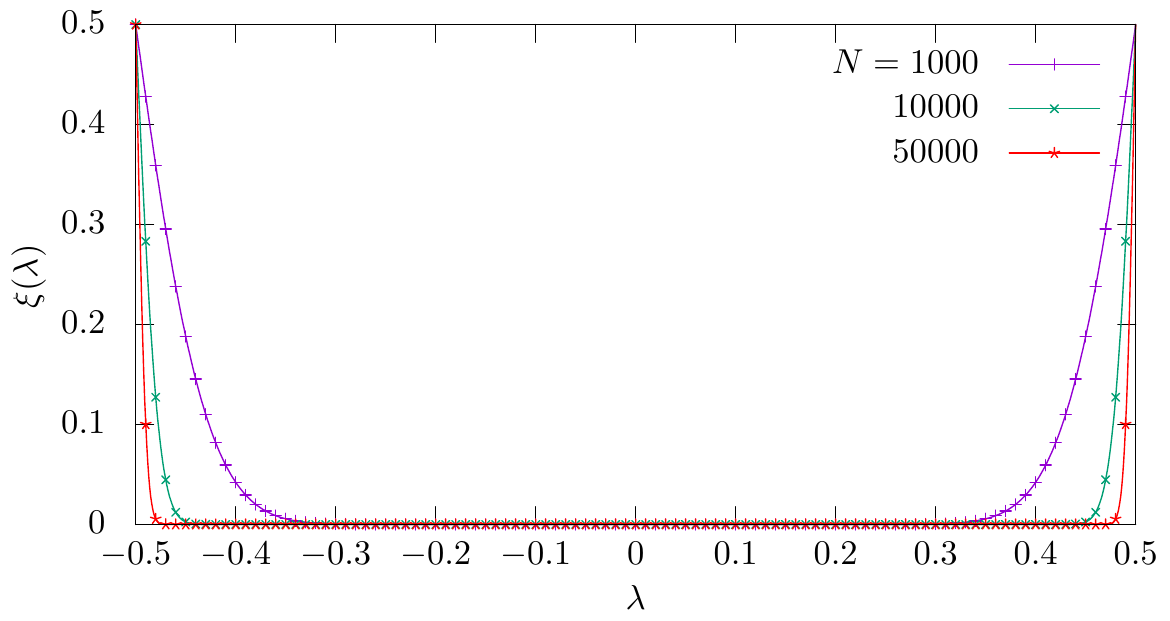}
  \caption{$\xi$ correction in the SSH model with periodic boundary
    condition. Using $1000$ ground states evenly sampled between
    $-0.5 < \lambda < 0.5$.  $\xi\sim 0.5$ close to the parameter
    boundaries $\lambda = \pm 0.5$, but quickly drops to zero away
    from them. With increasing system size $N$ ($2N$ lattice sites),
    the region of non-vanishing $\xi$ decreases. }
  \label{fig:ssh-erfc}
\end{figure}

We use the approximation
\begin{gather}
  \label{sm-cmn}
  C_{mn} \equiv \left|\langle \Psi_m |\Psi_n\rangle \right|^2 \simeq \exp \left[
    -g_m (n-m)^2 \delta\lambda^2 \right]\ ,
\end{gather}
where $g_m$ is the Fubini-Study metric,
\begin{gather}
  g_m \equiv g(\lambda_m) =
  \langle \partial_{\lambda} \Psi(\lambda_m) | \partial_{\lambda}
  \Psi(\lambda_m)\rangle - \langle \partial_{\lambda}\Psi(\lambda_m) |
  \Psi(\lambda_m) \rangle\langle \Psi(\lambda_m) | \partial_{\lambda}
  \Psi(\lambda_m)\rangle \ .  
\end{gather}
Then
\begin{align}
  \sum_{n=1}^M C_{mn} &= \sum_{x = 1-m}^{M-m} e^{-g_m x^2\delta\lambda^2} \longrightarrow \int\limits_{1-m}^{M-m} dx\ e^{-g_m x^2 \delta\lambda^2} \\
                      &= \int\limits_{-\infty}^{\infty} - \int\limits_{-\infty}^{1-m} - \int\limits_{M-m}^{\infty} \cdots = \frac{\sqrt{\pi}}{\sqrt{g_m}\delta\lambda} \left[ 1-\xi_m \right]\ ,
\end{align}
where the prefactor in the last expression results from the
infinite-limit Gaussian integral, and $\xi_m$ accounts for the second
and third integration ranges on the second line, \emph{i.e.}, the
effect of finite integration limits,
\begin{gather}
  \xi_m \equiv \xi(\lambda_m) = \frac{1}{2} \erfc\left[ \sqrt{g_m}\delta\lambda (m-1) \right] + \frac{1}{2} \erfc\left[ \sqrt{g_m} \delta\lambda (M-m) \right]\ , \\
  \erfc(x) = \frac{2}{\sqrt{\pi}}\int\limits_x^{\infty} dt\,e^{-t^2}\ .
\end{gather}
Note that the arguments in the $\erfc$s can be interpreted as
``quantum distances'' from $\lambda_m$ to the two boundaries
$\lambda_1$ and $\lambda_M$, respectively, as propagated by the metric
at $\lambda_m$. $\xi(\lambda) \simeq \frac{1}{2}$ at the boundaries
$\lambda_{a,b}$, but quickly approaches zero away from them,
increasingly so with larger system size. Fig.~\ref{fig:ssh-erfc} shows
its behavior in the SSH model with three different system sizes.

Upon normalization, this gives the weight
\begin{gather}
  \wt w_m = \frac{1}{M} \sum_{n=1}^M C_{mn} = \sqrt{\frac{\pi}{g_m}} \, \frac{1-\xi_m}{\lambda_M - \lambda_1} \, \frac{M-1}{M} \ .
\end{gather}
In the limit $M \rightarrow \infty$, the last fraction goes to $1$,
and we recover the continuum limit obtained in the main text.

$\dimp$ is defined as $1/\langle \wt w_m\rangle$. To proceed, we introduce
the length element and its relative fluctuation,
\newcommand{\dl}{\delta\ell}
\begin{gather}
  \dl_m \equiv \sqrt{g_m} \delta\lambda = \langle\dl_m\rangle(1 +
  \sigma_i) \quad , \quad \sigma_m \equiv \frac{\dl_m - \langle
    \dl_m\rangle}{\langle\dl_m\rangle} = \frac{M \dl_m}{\mathcal{L}} -
  1\ ,
\end{gather}
where $\mathcal{L} = \sum_{m=1}^M \dl_m = M \langle
\dl_m\rangle$. Then
\begin{align}
  \langle \wt w_m\rangle &= \frac{\sqrt{\pi}}{M} \langle
                           \frac{1-\xi_m}{\dl_m}\rangle = \frac{\sqrt{\pi}}{\mathcal{L}}
                           \langle \frac{1-\xi_m}{1+\sigma_m} \rangle \\
                         &= \frac{\sqrt{\pi}}{\mathcal{L}} \sum_{a=0}^{\infty}(-)^a \langle (1-\xi_m) \sigma_m^a\rangle \\
  \label{wm-exp}
                         &= \frac{\sqrt{\pi}}{\mathcal{L}} \left[ 
                           1 - \langle \xi_m\rangle + \langle \sigma_m^2\rangle + \langle \xi_m\sigma_m\rangle - \langle \xi_m\sigma_m^2\rangle + \mathcal{O}((1-\xi)\sigma^3)
                           \right]\ .
\end{align}
For a smooth curve, $\sigma_m^2 \ll 1$. Thus one can truncate at the
$\sigma_m^2$ order,
\begin{gather}
  \dimp = \frac{1}{\langle \wt w_m\rangle}  = \frac{\mathcal{L}}{\sqrt{\pi}} \langle \frac{1 -\xi_m}{1 + \sigma_m}\rangle^{-1} \simeq
  \frac{\mathcal{L}}{\sqrt{\pi}} (1 + \langle\xi_m\rangle - \langle
  \sigma_m^2\rangle + \cdots)\ .
\end{gather}
As discussed before, with increasing system size,
$\langle\xi_m\rangle$ decreases, thus in the thermodynamic limit, the
leading order correction should be $\langle\sigma_m^2\rangle$.

\begin{figure}
  \centering
  \includegraphics[width=.5\textwidth]
  {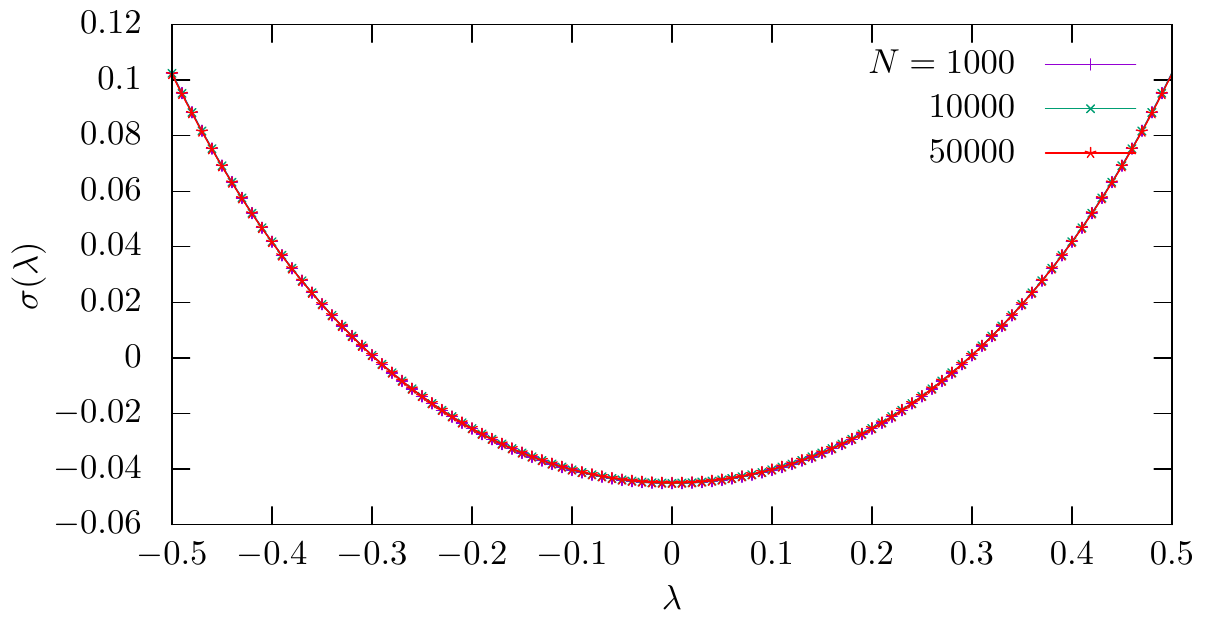}
  \caption{$\sigma$ correction in the SSH model with periodic boundary
    condition. Using $1000$ ground states evenly sampled between
    $-0.5 < \lambda < 0.5$.  Note that $\sigma$ is size independent
    ($2N = $ number of lattice sites).}
  \label{fig:ssh-sigma}
\end{figure}

Fig.~\ref{fig:ssh-sigma} shows $\sigma_m$ in the SSH model with three
different system sizes, note that it is size independent. Numerical
values of the expansion terms in Eq.~\ref{wm-exp} with $N=50000$ unit
cells are
\begin{gather}
  \begin{tabular}{c|c|c|c}
    $\langle \xi_m\rangle$ & $\langle \sigma_m^2\rangle$ & $\langle \xi_m\sigma_m\rangle$ & $\langle \xi_m\sigma_m^2\rangle$ \\
    \hline
    0.0067 & 0.0018 & 0.00067 & 0.000066
  \end{tabular}\ .
\end{gather}

\subsection{Alternative definition of characteristic dimensionality}
As discussed in the text, $1/\wt w_m$ counts the number of random states
in $\{|\Psi_n\rangle\}$ needed to represent $|\Psi_m\rangle$, and to
remove the $m$ dependence, there are two natural choices,
$\langle 1/\wt w_m\rangle$ and $1/\langle \wt w_m\rangle$, where
$\langle \cdots \rangle = \sum_{m=1}^M(\cdots) / M$. We have defined
$\dimp = 1 / \langle \wt w_m\rangle$ in the text, but the second choice
has its own merit, as it naturally evaluates to the geometric length.
In the thermodynamic limit, one can ignore the edge correction $\xi$
(see Fig.~\ref{fig:ssh-erfc}), then
\begin{gather}
  D_{alt} \equiv \langle \frac{1}{\wt w_m}\rangle = \frac{1}{M}\sum_{m=1}^M \frac{1}{\wt w_m} \stackrel{\xi_m = 0}{\longrightarrow}  \frac{1}{\sqrt{\pi}} \sum_{m=1}^M \sqrt{g_m}\delta\lambda = \frac{\mathcal{L}}{\sqrt{\pi}}\ .
\end{gather}

\end{document}